\documentclass[referee]{aa}
\usepackage{graphicx}
\usepackage{epsfig}

\def\be{\begin{equation}}
\def\ee{\end{equation}}
\def\bdm{\begin{displaymath}}
\def\edm{\end{displaymath}}

\def\cs {c_\mathrm{s}}
\def\va {v_\mathrm{a}}

\def\sn {_\mathrm{n}}

\newcommand{\vect}[1]{\vec{#1}}

\def\etac{\eta_\mathrm{C}}
\def\va {v_\mathrm{a}}

\begin{document}

\title
{The spatial damping of magnetohydrodynamic waves in a flowing partially ionised 
prominence plasma}

\author
{ M. Carbonell$^1$ \and P.  Forteza$^2$ \and R. Oliver$^2$ \and J. L. 
Ballester$^2$}

\offprints{J. L. Ballester\\ \email{joseluis.ballester@uib.es}}

\institute {Departament de Matem\`atiques i Inform\`atica, Universitat de les Illes Balears, 
E-07122 Palma de Mallorca, Spain
\and 
 Departament de F\'{\i}sica, Universitat de les Illes Balears, 
E-07122 Palma de Mallorca, Spain \\
\email{marc.carbonell@uib.es; pepfortezaferrer@gmail.com;
ramon.oliver@uib.es; joseluis.ballester@uib.es}}

\date{Received / Accepted}

\abstract
{Solar prominences are partially ionised plasmas displaying flows and oscillations. These oscillations show time and spatial 
damping and, commonly, have been 
explained in terms of magnetohydrodynamic (MHD) waves.}
{We study the spatial damping of linear non-adiabatic MHD waves in a flowing partially ionised plasma, having 
prominence-like physical properties. }
{We consider single fluid equations for a partially ionised hydrogen plasma including in the energy equation optically thin 
radiation, thermal conduction by electrons and neutrals, and heating. Keeping $\omega$ real and fixed, we have solved the 
obtained dispersion relations for the complex wavenumber, $k$, and have analysed the behaviour of the damping length, 
wavelength and the ratio of the damping length to the wavelength, versus period, for Alfv\'en, fast, slow and 
thermal waves.}
{The results indicate that in presence of a background flow new strongly damped fast and Alfv\'en waves appear which 
depend on the joint action of flow and resistivity. The damping lengths of adiabatic fast and slow waves are strongly affected 
by partial ionisation, which also modifies the ratio between damping lengths and 
wavelengths. Moreover, the behaviour of adiabatic fast waves looks 
very similar to that of Alfv\'en waves. For non-adiabatic slow waves, the unfolding in wavelength and damping length induced 
by the flow, allows that an efficient damping can be found for periods compatible with those 
observed in prominence oscillations. This effect is enhanced when low ionised plasmas are considered.}
{Since flows are ubiquitous in prominences, in the case of non-adiabatic slow 
waves and within the range of periods of interest for prominence oscillations, the joint effect of both flow and partial 
ionisation leads to a ratio between damping length and wavelength denoting a very efficient spatial damping. For fast 
and Alfv\'en waves, the most efficient damping occurs at very short periods not compatible with those observed in prominence 
oscillations.}
\titlerunning{Spatial damping of MHD waves in partially 
ionised plasmas} 
\authorrunning{Carbonell et al.} 
\maketitle

%%%%%%%%%%%%%%%%%%%%%%%%%%%%%%%%%%%%l
\section{Introduction}\label{intro}
%%%%%%%%%%%%%%%%%%%%%%%%%%%%%%%%%%%%

A solar prominence is a cool ($T \sim 10^{4}$\ K) and dense ($\rho \sim 5 \cdot 10^{11}$ \ kg/m$^{3}$) mass of gas located in
the much denser and hotter solar corona.  Although it is not well understood how this structure can live for a long time within
the solar corona, it is thought that its support and thermal isolation are of magnetic origin.  Thanks to ground and
space-based observations, evidence about the presence of small-amplitude oscillations in prominences and filaments is now
widely available and detailed information can be found in Oliver and Ballester (2002), Banerjee et al.  (2007), Oliver,
(2009), and Mackay et al.  (2009).  The oscillations are mainly detected through the periodic Doppler shifts of
spectral lines, with periods from a few minutes to hours, and the observations have shown that they are of local
nature, that they can be coherent over large regions of prominences/filaments, that simultaneously flowing and oscillating
features are present, and that oscillations display time and spatial damping (Terradas et al. 2002).  These small-amplitude oscillations have been
commonly interpreted in terms of linear magnetohydrodynamic (MHD) waves and the damping of oscillations has been studied by
considering different dissipative mechanisms.  For instance, in the case of fully ionised plasmas the time damping of
prominence oscillations has been studied by considering non-adiabatic MHD waves, including the effect of optically thin
radiation, thermal conduction and heating, in unbounded and bounded prominence plasmas (Carbonell et al.  2004; Terradas et
al.  2005); furthermore, assuming that the threads that seem to compose solar filaments can be modeled as cylindrical flux
tubes, the time damping of homogeneous and inhomogeneous cylindrical flux tubes, with prominence physical conditions, has
been modeled using thermal mechanisms and resonant absorption, respectively (Soler et al.  2008; Arregui et al.  2008; Soler
et al.  2009b).

A typical feature in prominence oscillations is the presence of flows which are observed in $H\alpha$, UV and EUV lines 
(Labrosse et al. 2009).  In
$H\alpha$ quiescent filaments, the observed velocities range from $5$ to $20$ km/s (Zirker et al.  1998; Lin et al.  2003,
2007) and, because of physical conditions in filament plasma, they seem to be field-aligned.  In the case of active region
prominences, flow speeds can be higher.  Recently, observations made with Hinode/SOT by Okamoto et al.  (2007) reported the
presence of synchronous vertical oscillatory motions in the threads of an active region prominence, together with the
presence of flows along the same threads.  However, in limb prominences different kinds of flows are observed and, for
instance, observations made by Berger et al. (2008) with Hinode/SOT have revealed a complex dynamics with vertical downflows
and upflows.  Taking all this into account, Carbonell et al (2009) explored the time damping of non-adiabatic slow and
thermal waves in an unbounded prominence medium with a background flow, while Soler et al.  (2008, 2009a) investigated the
time damping of the oscillations of an individual prominence thread and of a threaded prominence when both mass flows and
non-adiabatic processes are considered.

Although there is a large amount of observational evidence about the time damping of MHD waves propagating in coronal
structures such as coronal loops or prominences, current observational information about the spatial damping of MHD waves in
coronal structures is still scarce.  Time damping is usually produced when an impulsive perturbation excites a medium and
waves start to propagate, then, the wave amplitude decreases as the time goes by.  Alternatively, if the medium is excited
by a continuous driver with a fixed frequency, the spatial damping is detected by the decrease of the amplitude when the wave
propagates.  Since, probably, most of the phenomena exciting waves in solar structures are of impulsive origin rather than
produced by continuous drivers, time damping of waves is more often observed.

 Concerning spatial damping, De Moortel et al.  (2002) and De Moortel \& Hood (2003, 2004), using thermal conduction,
compressive viscosity, gravitational stratification and field line divergence, studied the spatial damping of driven and
non-driven slow MHD waves in coronal conditions, applying the obtained results to the case of standing and propagating slow
waves in coronal loops observed with SOHO and TRACE. In the case of prominences, Terradas et al.  (2002) analyzed small
amplitude oscillations in a polar crown prominence reporting the presence of a plane propagating wave as well as an standing
wave.  The plane wave propagates in opposite directions with wavelengths of $67,500$ and $50,000$ km and phase speeds of $15$
km~s$^{-1}$ and $12$ km~s$^{-1}$, while in the case of the standing wave, the estimated wavelength is of $44,000$ km and the
phase speed of $12$ km~s$^{-1}$.  These authors also reported that in the case of the propagating wave, which was interpreted
as a slow MHD wave, the amplitude of the oscillations spatially decreases in a substantial way in a distance of $2-5 \times
10^4$ km from the location where wave motion was being generated.  This distance can be considered as a typical spatial
damping length, $L_\mathrm{d}$, of the oscillations.

From the theoretical point of view, and assuming a fully ionised plasma, Ballai (2003) performed a qualitative study of the
spatial damping of linear compressional waves in solar prominences by considering different dissipative mechanisms such as
isotropic and anisotropic viscosity, isotropic magnetic diffusivity, isotropic radiative damping (Newton's law) and
anisotropic thermal conduction.  The conclusion was that thermal radiation can damp compressional waves as well as that waves
can be also damped by anisotropic thermal conduction provided that they have enough short wavelengths.  Using linear
non-adiabatic MHD waves, which included optically thin radiation, thermal conduction and heating, Carbonell et al.  (2006)
made a quantitative study of the spatial damping of slow, fast and thermal waves in a fully ionised unbounded prominence
medium.  In the frequency space, they determined the regions where radiation or thermal conduction are the dominant damping
mechanisms, the critical frequencies at which the dominance changes from one mechanism to another, and the fact that
different heating mechanisms do not strongly affect the damping.  The most important conclusions were that the thermal wave
is propagating but strongly damped, and that slow waves are efficiently damped by thermal effects, which does not happen for
fast waves.  Singh et al.  (2006) did a similar study but considering only Newtonian radiation and, later, Singh et al.
(2007) considered again the same problem including Newtonian radiation and turbulent viscosity.

The typical composition of solar prominences is $90\%$ hydrogen and $10\%$ helium and they constitute partially ionised
plasmas since hydrogen lines are observed (Labrosse et al. 2009).  The exact ionisation degree of prominences is unknown and the ratio of electron
density to neutral hydrogen density seems to be in the interval $0.1 - 10$ (Patsourakos and Vial, 2002).  From a theoretical
point of view, partial ionisation was already considered by Mercier and Heyvaerts (1977) when they studied the difussion of
neutral atoms due to gravity; also, Gilbert et al.  (2002) studied the diffusion of neutral atoms in a partially ionized
prominence plasma, concluding that the loss time scale is much longer for hydrogen than for helium.  Recently, Gilbert et al.
(2007) have investigated the temporal and spatial variations of the relative abundance of helium with respect to hydrogen in
a sample of filaments.  They have found that a majority of filaments show a deficit of helium in the top part while in the
bottom part there is an excess.  This seems to be due to the shorter loss time scale of neutral helium compared to that of neutral
hydrogen.  The consideration of prominences as made of partially ionised plasma is extremely important for the physics of
prominences, and the effects on MHD waves in prominences need to be taken into account.  From the theoretical point of view,
and in the framework of laboratory plasma physics, a extense literature about wave propagation in a partially ionized
multifluid plasma can be found (Watanabe, 1961a,b; Tanenbaum, 1961; Tanenbaum and Mintzer, 1962; Woods, 1962; Kulsrud and
Pearce, 1969; Watts and Hanna, 2004).  In astrophysical plasmas, the typical frequency of MHD waves is much smaller than the
collisional frequencies between species.  In such a case the single fluid approach is usually adopted and it has been applied
to wave damping in the solar atmosphere (De Pontieu, et al.  2001; Khodachenko et al.  2004; Leake et al.  2005).  In the
case of solar prominences, Forteza et al.  (2007) derived the full set of MHD equations for a partially ionized, single fluid
plasma and applied them to study the time damping of linear, adiabatic waves in an unbounded prominence medium.  Later, this
study was extended to the non-adiabatic case by including thermal conduction by neutrals and electrons, radiative losses and
heating (Forteza et al.  2008).  Because of the effect of neutrals, in particular that of ion-neutral collisions, a
generalized Ohm's law has to be considered, which causes some additional terms to appear in the resistive magnetic induction
equation in comparison to the fully ionized case.  Among these additional terms, the dominant one in the linear regime is the
so-called ambipolar magnetic diffusion, which enhances magnetic diffusion across magnetic field lines.  Furthermore, one of
the interesting effects produced by the consideration of partial ionisation and ion-neutral collisions is that Alfv\'en waves
can be damped, which cannot be obtained by means of thermal mechanisms, and Singh \& Krishnan (2009) have studied the time
damping of Alfv\'en-like waves in a partially ionised plasma representing a particular model of the solar atmosphere.  For
bounded media, Soler et al.  (2009c) have applied this formalism to the study of the time damping of fast, Alfv\'en and slow
waves in a partially ionised filament thread modeled as a cylinder, and Soler et al.  (2009d) have used a cylindrical filament
thread, having an inhomogeneous transition layer between prominence and coronal material, to study the influence of partial
ionisation on the time damping of fast kink waves which is caused by resonant absorption in the inhomogeneous layer.  

Up to now, and to our knowledge, no study of the spatial damping of MHD waves in a partially ionised plasma has been
performed.  Therefore, since solar prominences are partially ionised plasmas in which material flows and oscillations are
present and since commonly these oscillations are interpreted in terms of MHD waves, our main aim here is to explore the
theoretical and observational effects associated to the spatial damping of non-adiabatic MHD waves in an unbounded and
partially ionised plasma, with prominence-like physical conditions, when a background flow is present.  The layout of the
paper is as follows: In Sect.  2, the equilibrium model and some theoretical considerations are presented; in Sect.  3, the
spatial damping of Alfv\'en waves is studied; in Sect.  4, the spatial damping of magnetoacoustic waves is considered;
finally, in Sect.  5, conclusions are drawn.

%%%%%%%%%%%%%%%%%%%%%%%%%%
\section{Model and Methods} \label{mod}
%%%%%%%%%%%%%%%%%%%%%%%%%%%

%%%%%%%%%%%%%%%%%%%%%%%%%
\subsection{Equilibrium Properties} \label{eq}
%%%%%%%%%%%%%%%%%%%%%%%%%%

As a background model, we use a homogeneous unbounded medium threaded
by a uniform magnetic field along the $x$-direction, and with a
field-aligned background flow.  The equilibrium magnitudes of the
medium are given by
$$
  p_0 =\mathrm{const.}, \ \ \rho_0 = \mathrm{const.}, \ \ T_0 = \mathrm{const.},
  $$
  $${\bf B_0} = 
  B_{0}\hat x, \ \ {\bf v}_0 =  v_{0}\hat x,
$$
with $B_{0}$ and $v_{0}$ constants, and the effect of gravity has 
been ignored. Since we consider a medium with physical properties akin to those of a 
solar prominence, the density is $\rho_\mathrm{0} = 5 
\times 10^{-11}$  kg/m$^{3}$, the  temperature $T_\mathrm{0} = 8000$ \ K, the 
magnetic field $\vert \bf B_\mathrm{0} \vert$= 10 \ G, and, in general, a field-aligned flow with $v_\mathrm{0} = 10$ km/s, simulating the typical 
flows found in quiescent filaments, has been considered.

%%%%%%%%%%%%%%%%%%%%%%%%%
\subsection{Basic and Linearised Equations} \label{le}
%%%%%%%%%%%%%%%%%%%%%%%%%

The derivation of the single-fluid MHD equations for a partially ionised hydrogen plasma can be found in Forteza et al.
(2007).  Later, these equations were modified in order to include non-adiabatic processes in the energy equation (Forteza et
al.  2008), and the physical meaning of all the terms and quantities used in the following can be found in those papers.
When a background flow is considered, the single fluid basic equations for the study of non-adiabatic MHD waves in a 
partially ionised plasma can be derived from Forteza et al.  (2007, 2008) and Carbonell et al.  (2009).  The most important difference with
respect to the non-adiabatic case without a background flow is that instead of the operator $\frac{\partial}{\partial t}$ we
will have $\frac{\partial}{\partial t} + {\bf v}_{0} \cdot \nabla$ in the MHD equations (Goedbloed and Poedts, 2004).
 Before proceeding, some remarks are in order: We consider a partially ionised hydrogen plasma characterized by a plasma density,
$\rho_0$, temperature, $T_0$, and number densities of neutrals, $n_{n}$, ions, $n_{i}$, and electrons $n_{e}$, with
$n_{e}=n_{i}$.  Thus, the gas pressure is $p_0 = (2 n_{i} + n_{n}) k_{\rm B} T_0$, where $k_{ B}$ is Boltzmann's constant.
The relative densities of neutrals, $\xi_{n}$, and ions, $\xi_{i}$, are given by
\begin{equation}
 \xi_{n} = \frac{n_{n}}{n_{i} + n_{n}}, \qquad \xi_{i} = \frac{n_{i}}{n_{i} + n_{n}},
\end{equation}
where we have neglected the contribution of electrons.  We can now define a ionization fraction, $\tilde \mu$, which gives us information
about the plasma degree of ionization,
\begin{equation}
 \tilde{\mu} = \frac{1}{1+\xi_{i}}.
\end{equation}
and the quantity
\begin{equation}
 \Xi = \frac{ \xi_{n}^{2}\xi_{i}}{(1+\xi_{i}) \alpha_{n}}.
\end{equation}
where $\alpha_{n}$ is a friction coefficient (Braginskii, 1965; Khodachenko et al. 2004; Leake et al. 2005).
For a fully ionised plasma $\tilde{\mu} = 0.5$ while for a neutral plasma $\tilde{\mu} = 1$.  Depending on the value given to
$\tilde \mu$ and to Spitzer's ($\eta$) and Cowling's ($\eta_\mathrm{C}$) resistivities, we may have different types of plasmas: A
fully ionised ideal plasma (FIIP) is characterised by $\tilde \mu = 0.5$, $\eta = \eta_\mathrm{C} = \Xi = 0$; a fully ionised
resistive plasma (FIRP) is characterised by $\tilde \mu = 0.5$, $\eta = \eta_\mathrm{C}$ and $\Xi = 0$; and a partially ionised
plasma (PIP) is characterised by $\tilde 0.5 < \tilde \mu < 1$, $\eta \neq \eta_\mathrm{C}$ and $\Xi \neq 0$.  Although our main aim
is the study of spatial damping of MHD waves in PIP, the case of FIRP will be also considered since it has received
almost no attention in the literature.  Another important issue for our study is the numerical value and behaviour of the sound
$(c_\mathrm{s})$ and Alfv\'en speeds $(v_\mathrm{a})$.  For fixed the density and magnetic field, the Alfv\'en speed has a constant
numerical value of $126.15$ \ km/s, however, since the sound speed depends on gas pressure, which is a function of the number
densities of ions and neutrals, its numerical value is not constant but depends on the ionisation fraction considered.  For a
fully ionised plasma, the sound speed is $14.84$ \ km/s while for a partially ionised plasma with $\tilde \mu = 0.95$ its
value decreases to $10.76$ \ km/s.  This variation in the sound speed can be important since depending on the flow speed and
ionisation fraction chosen, the flow speed could be greater than, smaller than or equal to the sound speed which affects the
direction of propagation of slow and thermal waves (Carbonell et al.  2009).  In our case, the flow speed is subsonic and
subalfv\'enic which seems to be a typical feature of flows observed in quiescent filaments.

Next, to obtain the dispersion
relation for linear MHD waves, we consider small perturbations from the equilibrium of the form
\begin{eqnarray}& &{}
\vect{B}(t,\vect{r})=\vect{B}_0+\vect{B}_1(t,\vect{r}),\qquad 
p(t,\vect{r})=p_0+p_1(t,\vect{r}),\nonumber{}\\ & &{}
\rho(t,\vect{r})=\rho_0+\rho_1(t,\vect{r}), \qquad 
T(t,\vect{r})=T_0+T_1(t,\vect{r}), \qquad \nonumber{} \\ & &{}
\vect{v}(t,\vect{r})= \vect{v}_0+\vect{v}_1(t,\vect{r}).\nonumber{}
\end{eqnarray}
\noindent 
and we linearise the single fluid basic equations. Since the medium is unbounded, we perform a Fourier analysis  in terms of plane waves and assume that perturbations behave as
\be f_1(\vect{r},t)=fe^{i(\omega t-\vect{k}\cdot\vect{r})}.\ee
\noindent and with no loss of generality, we choose the $z$-axis so 
that the wavevector $\vect{k}$ lies in the $xz$-plane 
($\vect{k}=k_x\vect{\hat{x}}+k_z\vect{\hat{z}}$), and introducing the propagation angle, $\theta$, between $\vect{k}$ and
$\vect{B}_0$, the wavenumber components can be expressed as $k_x=k\cos\theta$ and $k_z=k\sin\theta$.  With this approach, the operator $\frac{\partial}{\partial t} + {\bf v}_{0} \cdot \nabla$ becomes $i(\omega - k_{x} v_{0})$, which points out that in the presence of a
background flow the frequency has a
Doppler shift given by $k_{x} v_{0}$ and that the wave frequency, $\omega$,
for the non-adiabatic case with flow can be obtained from
\begin{equation} \label{freq}
\omega = \Omega + k_{x} v_{0}, \label{om}
\end{equation}
$\Omega$ being the wave frequency for the non-adiabatic case without 
flow. Also, these frequencies can be described in a 
different manner, $\Omega$ corresponds to the frequency measured by 
an observer linked to the flow inertial rest frame, while $\omega$ would 
correspond to the frequency measured by an observer linked to an 
external inertial rest frame.

After the Fourier analysis, the following linearised scalar equations are obtained.

\be \Omega\rho_1 - \rho_0(k_xv_{1x}+k_zv_{1z})=0,\label{cont_m}\ee
\be \Omega\rho_0v_{1x}-k_xp_1=0,\label{mom_x}\ee
\be \Omega\rho_0v_{1y}+\frac{B_{0x}}{\mu}k_xB_{1y}=0,\label{mom_y}\ee
\be \Omega\rho_0v_{1z}-k_zp_1+\frac{B_{0x}}{\mu}(k_xB_{1z}-k_zB_{1x})=0 \label{mom_z},\ee
\begin{eqnarray} i\Omega 
    \left(p_1-\cs^2\rho_1\right)+(\gamma-1)\left(\kappa_{\mathrm{e}\parallel}k_x ^2+\kappa\sn k^2 +\rho_0L_T\right)T_1+\nonumber{}\\
+(\gamma-1)(L+\rho_0L_\rho)\rho_1=0,\label{ener}\end{eqnarray}
\begin{eqnarray} 
    B_{1x}\left(i\Omega+k_x^2\eta+k_z^2\etac\right)+(\eta-\etac)k_xk_zB_{1z}-\nonumber{}\\-B_{0x}k_z(iv_{1z}-k_z\Xi p_1)=0,\label{ind_x}\end{eqnarray}
\be B_{1y}\left(i\Omega+k_x^2\etac+k_z^2\eta\right)+iB_{0x}k_xv_{1y}=0,\label{ind_y}\ee
\begin{eqnarray} 
    B_{1z}\left(i\Omega+k_x^2\etac+k_z^2\eta\right)+(\eta-\etac)k_xk_zB_{1x}+\nonumber{}\\+B_{0x}k_x(iv_{1z}-k_z\Xi p_1)=0,\label{ind_z}\end{eqnarray}
\be k_xB_{1x}+k_zB_{1z}=0,\label{div_b}\ee
\be \frac{p_1}{p_0}-\frac{\rho_1}{\rho_0}-\frac{T_1}{T_0}=0.\label{stat}\ee

%%%%%%%%%%%%%%%%%%%%%%%%%%%%%
\subsection{Dispersion relation for Alfv\'en waves}
%%%%%%%%%%%%%%%%%%%%%%%%%%%%%%

\noindent Equations~(\ref{mom_y}) and (\ref{ind_y}) are decoupled 
from the rest and from them we can obtain a dispersion relation for 
Alfv\'en waves in a partially ionised plasma with a background flow, which is,
\begin{eqnarray}
    \Omega^{2} -i \Omega k^{2}( \eta_\mathrm{C} \cos^{2} \theta + \eta \sin^{2} \theta)-v_\mathrm{a}^{2} k^{2} \cos^{2} \theta = 0
 \label{disp_alf1}
\end{eqnarray}
where $\theta$ is the angle between the wavenumber vector and the magnetic field.
 In absence of flow, Eq.~(\ref{disp_alf1}) reduces to,

\be k^{2} = \frac{\omega^2}{ \Gamma(\theta)^2 \cos^{2} \theta} ,\label{disp_alf2}\ee
\noindent 
where $\Gamma(\theta)$ is a modified and complex Alfv\'en speed ($\Gamma(\theta)=\Gamma_{\mathrm{r}}+i\Gamma_{\mathrm{i}}$) defined as
\be \Gamma(\theta)^2=\va^2+i\omega\left(\etac+\eta\tan^2\theta\right).\label{va_mod}\ee
which was introduced by Forteza et al. (2008). In the case of a fully ionised ideal plasma without a background flow Eq.~(\ref{disp_alf1}) reduces to the 
well known expression for Alfv\'en waves.

%%%%%%%%%%%%%%%%%%%%%%%%%%%%%%%%
\subsection{Dispersion relation for magnetoacoustic waves}
%%%%%%%%%%%%%%%%%%%%%%%%%%%%%%%%

From the remaining linearised equations, imposing that the determinant of the algebraic system be zero, we obtain our general
dispersion relation for thermal and magnetoacoustic waves in presence of a background flow, which is given by,
\begin{eqnarray}
    (\Omega^{2} -k^{2} \Lambda^{2}) (ik^{2} \eta_\mathrm{C} \Omega-\Omega^{2})+k^{2} v_\mathrm{a}^{2}(\Omega^{2} -k_{x}^{2} \Lambda^{2} 
    )+ \nonumber \\
    +i k^{2} k_{z}^{2}v_\mathrm{a}^{2} \Lambda^{2} \Xi \rho_{0} \Omega= 0
 \label{disp_mag}
\end{eqnarray}
where $\Lambda^{2}$ is the non-adiabatic sound speed squared (Forteza et al. 2008, Soler et al. 2008) and is defined as
\begin{eqnarray}
    \Lambda^{2} = \frac{\frac{T_{0}}{\rho_{0}} A - H +ic_\mathrm{s}^{2} \Omega}{\frac{T_{0}}{p_{0}}A+i \Omega} \label{nass}
    \end{eqnarray}
where $A$ and $H$, including optically thin radiative losses, thermal conduction by electrons and neutrals, and a constant
heating per unit volume, are defined in Forteza et al.  (2008).  When Eq.~(\ref{disp_mag}) is expanded, it becomes a seventh
degree polynomial in the wavenumber $k$.  By setting the appropiate values to the corresponding quantities, the dispersion
relation (\ref{disp_mag}) is consistent with the case of an adiabatic and partially ionised plasma without flow (Forteza et
al.  2007), the case of a non-adiabatic and partially ionised plasma without flow (Forteza et al.  2008) and the case of a
non-adiabatic, fully ionised plasma with a background flow (Carbonell et al.  2009).

%%%%%%%%%%%%%%%%%%%%%%%%%%%%%%%
\section{Spatial damping of Alfv\'en waves in a partially ionised 
    plasma}  
%%%%%%%%%%%%%%%%%%%%%%%%%%%%%%%%
Since we are interested in the spatial damping of MHD waves we consider the frequency, $\omega$, to be real
and seek for complex solutions of the wavenumber $k$ expressed as $k = k_\mathrm{r} + ik_\mathrm{i}$.  The wavelength of the waves is given
by $\lambda = \frac{2 \pi}{k_\mathrm{r}}$, the damping length by $L_\mathrm{d} = \frac{1}{k_\mathrm{i}}$ and the damping length per wavelength is
$\frac{L_\mathrm{d}}{\lambda}$ and, in general, we consider a propagation angle $\theta = \frac{\pi}{4}$.

  %%%%%%%%%%%%%%%%%%%%%%%%%%%%%%%%%%%%%%%%%%%%  
\subsection{Spatial damping of Alfv\'en waves without background flow} \label{Anf}
\begin{figure}
	  \centering{
		  \resizebox{9cm}{!} {\includegraphics{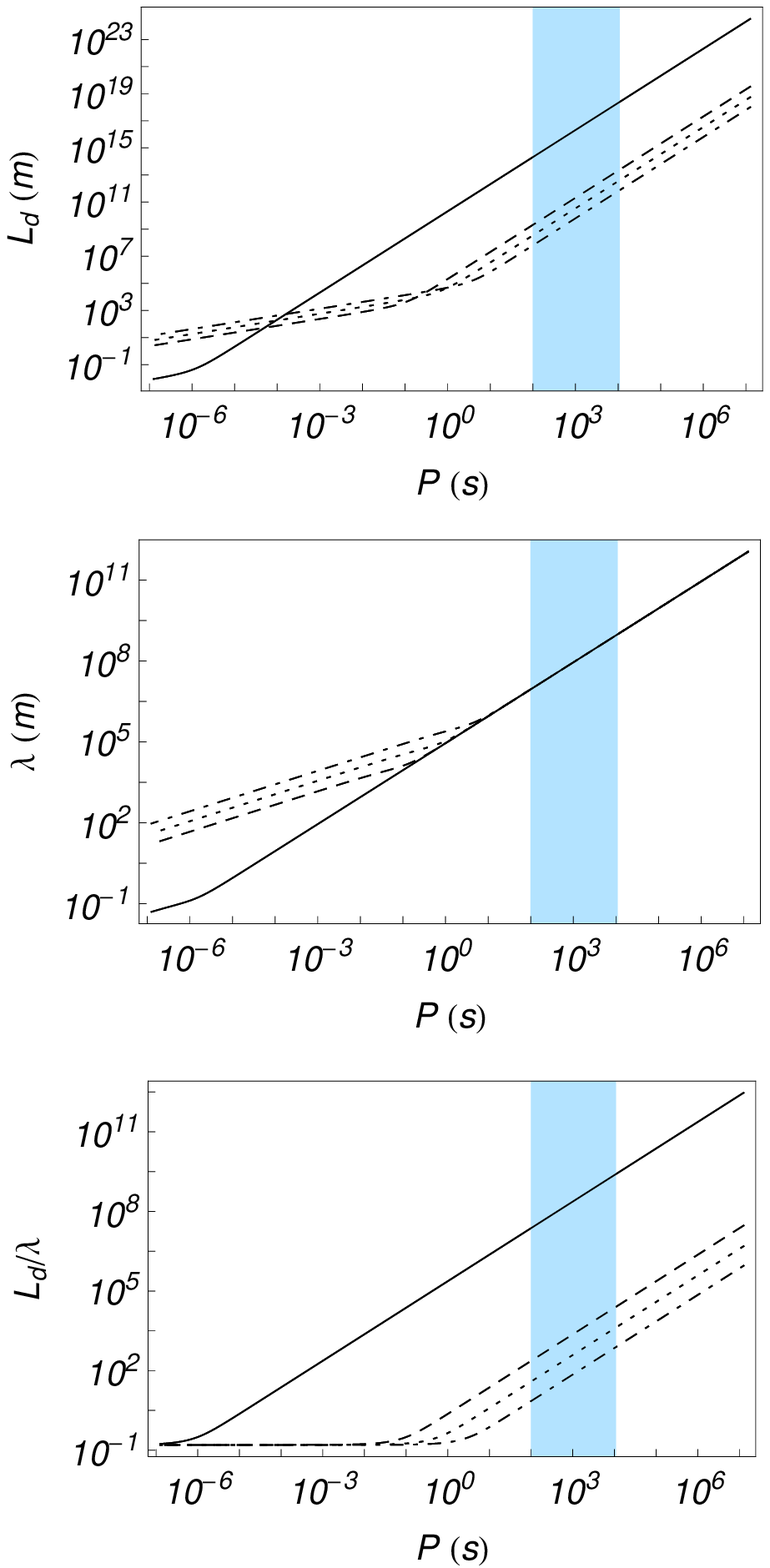}
		      \includegraphics{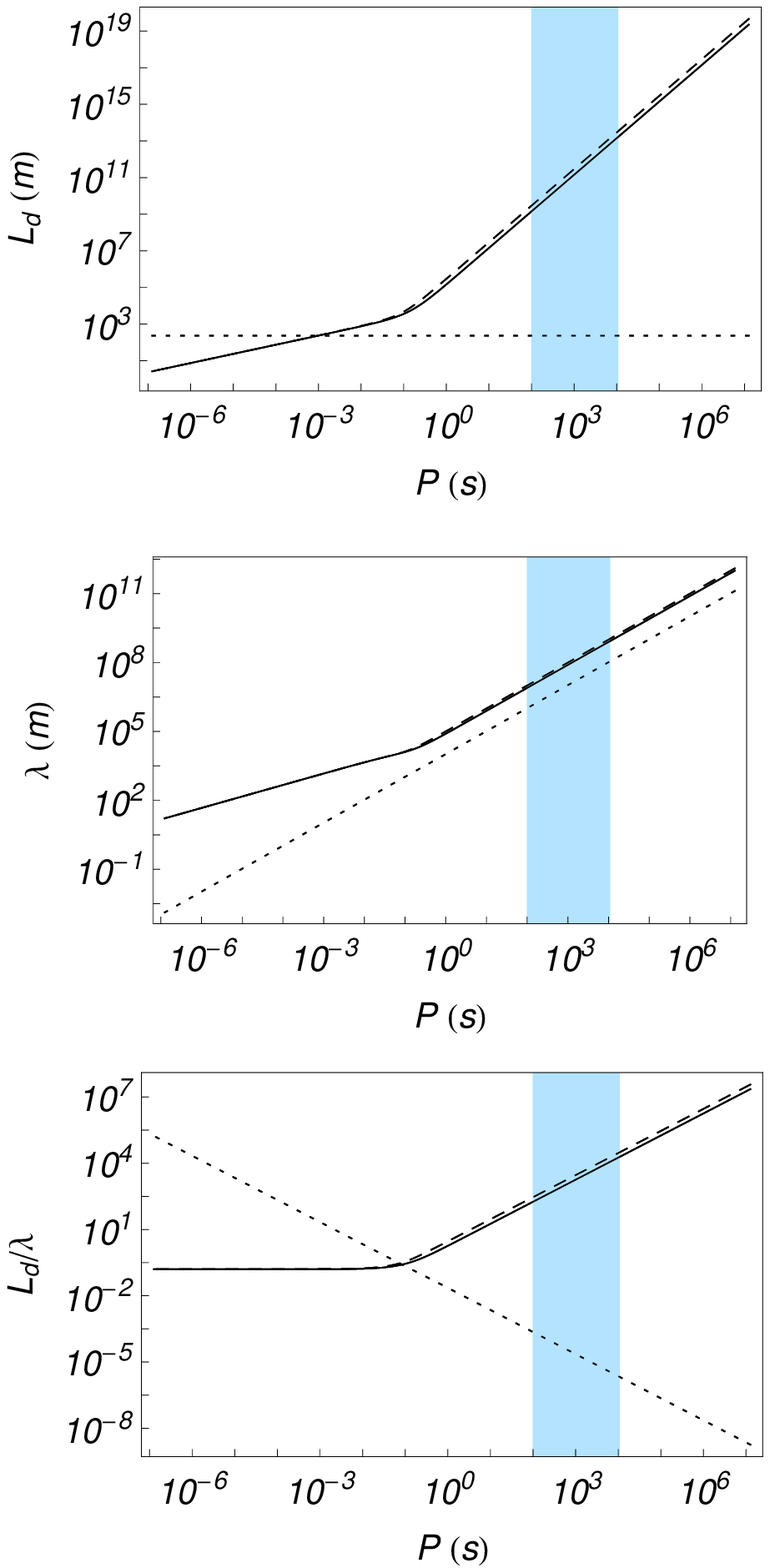}}}
		   \vspace{-3mm}
		   \caption{Left panels: Damping length, wavelength, and ratio of the damping length to the wavelength versus the period
for Alfv\'en waves in a FIRP (solid) and in PIP with
$\tilde{\mu}=0.8$ (dashed), $\tilde{\mu}=0.95$ (dotted), and $\tilde{\mu}=0.99$ (dash-dotted). Right panels:	   
Damping length, wavelength, and ratio of the damping length to the wavelength versus period
for the three (solid, dashed, dotted) Alfv\'en waves in a PIP with  
$\tilde{\mu}=0.8$ and with a background flow of $10$ \ km/s. In all the Figures, the shaded region 
corresponds to the interval of observed periods in prominence 
oscillations.}      \label{f01} 
	   \end{figure}

%%%%%%%%%%%%%%%%%%%%%%%%%%%%%%%%%%%%%%%%%%%%%
In this case, our governing dispersion relation is given 
by Eq.~(\ref{disp_alf2}), and the wavenumbers are
\be k = \pm \sqrt{\frac{(\omega
\sec \theta)^{2}}{v_\mathrm{a}^{2}+i\omega (\eta_\mathrm{C}+\eta \tan^{2} \theta)} }\ee 
representing two Alfv\'en waves propagating in opposite directions.
The real part ($k_{r}$) of these wavenumbers  is,
 \begin{eqnarray} 
k_\mathrm{r} = \omega \sec \theta \frac{ 
 \sqrt{v_\mathrm{a}^{2}+\sqrt{v_\mathrm{a}^{4} + \omega^{2} (\eta_\mathrm{C} +\eta \tan 
 ^{2} \theta)}}}{\sqrt{2 \left[v_\mathrm{a}^{4} + \omega^{2} (\eta_\mathrm{C} +\eta \tan 
 ^{2} \theta)\right]}} \label{kra}
 \end{eqnarray}
 while the imaginary part ($k_{i}$) is,
 \begin{eqnarray}
  k_\mathrm{i} & = &   \frac{-\omega^{2} \sec \theta}{\sqrt{2 \left[v_\mathrm{a}^{4} + \omega^{2} (\eta_\mathrm{C} +\eta \tan 
 ^{2} \theta)\right]}} \times  \nonumber 
 \end{eqnarray}
 \begin{eqnarray}
 \times \frac{(\eta_\mathrm{C}+\eta \tan^{2} 
 \theta)}{\sqrt{v_\mathrm{a}^{2}+\sqrt{v_\mathrm{a}^{4} + \omega^{2} (\eta_\mathrm{C} +\eta \tan 
 ^{2} \theta)}}}  \label{kia}
  \end{eqnarray}
Figure~\ref{f01} (Left panels) shows a plot of the damping length, wavelength, and the ratio of the damping length to wavelength,
versus the period, for Alfv\'en waves.  The plots have been made for four different ionisation fractions and the shaded region 
corresponds to the interval of observed periods in prominence 
oscillations. When a period
within the shaded region is considered, we observe that the damping length decreases in a substantial way when the amount of
neutrals in the plasma increases.  Then, when ion-neutral collisions are present the spatial damping of Alfv\'en waves is
enhanced for periods greater than $1$ s.  Also, we can observe that for a FIRP the behaviour of the damping length, 
wavelength and their ratio versus period is linear, except for periods below $10^{-6}$ s. However, when a PIP is considered a deviation from the linear behaviour appears for periods below $1$ s.  This is due to the joint effect of the terms including frequency and resistivities in the real $(k_{r})$ and imaginary $(k_{i})$ parts of the wavenumber $k$
since going to shorter periods frequency increases and its role becomes more important.  The spatial damping of Alfv\'en
waves in PIP with physical properties like those of quiescent prominences is very efficient for periods below $1$ s, however, within the interval of periods of interest for prominence oscillations, it is only efficient, ($\frac{L_\mathrm{d}}{\lambda} \sim 1 - 10$), when almost neutral plasmas are considered.

%%%%%%%%%%%%%%%%%%%%%%%%%%%%%%%%%%%%%
 \subsection{Spatial damping of Alfv\'en waves with background flow} 
 \label{Af}
 %%%%%%%%%%%%%%%%%%%%%%%%%%%%%%%%%%%%%
 
Now, our dispersion relation is given by Eq.~(\ref{disp_alf1}) which once expanded becomes a cubic polynomial in the wavenumber 
$k$,
such as,
\begin{eqnarray}
iv_\mathrm{0}(\etac \cos^{2} \theta + \eta  \sin^{2} 
\theta)k^{3}+ \nonumber \\
+ \left[(v_\mathrm{0}^{2}-v_\mathrm{a}^{2}-i \etac \omega)\cos \theta-i 
\eta \omega \sin \theta \tan \theta\right]k^{2} - 2 \omega 
v_\mathrm{0}k+\nonumber{}\\
+ \omega^{2} \sec \theta=0, \label{disp_alf5}
\end{eqnarray}
The increase in the degree of the dispersion relation, with respect to the case without flow, is produced by the joint presence of flow and resistivities.  In this
case, we should obtain three propagating Alfv\'en waves, therefore, Figure~\ref{f01} (Right panels) shows the numerical solution of dispersion relation (\ref{disp_alf5}) for the three Alfv\'en waves in a
partially ionised plasma with a background flow.   For all the interval of periods considered, a strongly damped third Alfv\'en wave appears, while on the contrary, and like in~\ref{Anf}, the other two Alfv\'en waves are very efficiently damped for periods
below $1$ s. However, within the interval of periods typically observed in prominence oscillations these waves are only efficiently attenuated 
 when almost neutral plasmas are considered.  Furthermore, the
following approximations for the different wavenumbers corresponding to two of the expected Alfv\'en waves can be calculated,
\be k \simeq \pm \sqrt{\frac{(\omega
\sec \theta)^{2}}{(v_\mathrm{0} \pm v_\mathrm{a})^{2}+i\omega (\eta_\mathrm{C}+\eta \tan^{2} \theta)} }\label{disp_alf3} \ee 
whose real part ($k_\mathrm{r}$) is given by, 
\begin{eqnarray} 
k_\mathrm{r} & \simeq &  \omega \sec \theta  \frac{ \sqrt{(v_\mathrm{0} \pm v_\mathrm{a})^{2}+\sqrt{(v_\mathrm{0} \pm v_\mathrm{a})^{4} + \omega^{2} 
(\eta_\mathrm{C}
+\eta \tan ^{2} \theta)}}}{\sqrt{2 \left[(v_\mathrm{0} \pm v_\mathrm{a})^{4} + \omega^{2} (\eta_\mathrm{C} +\eta \tan ^{2} \theta)\right]}}
\end{eqnarray}
while the imaginary part ($k_\mathrm{i}$) is, 
\begin{eqnarray} 
k_\mathrm{i} & \simeq & \frac{-\omega^{2} \sec \theta}{\sqrt{2 \left[(v_\mathrm{0} \pm v_\mathrm{a})^{4} + \omega^{2} 
(\eta_\mathrm{C} +\eta \tan ^{2} \theta)\right]}} \times \nonumber
 \end{eqnarray}
\begin{eqnarray} 
 \times  \frac{(\eta_\mathrm{C}+\eta \tan^{2}
\theta)}{\sqrt{(v_\mathrm{0} \pm v_\mathrm{a})^{2}+\sqrt{(v_\mathrm{0} \pm v_\mathrm{a})^{4} + \omega^{2} (\eta_\mathrm{C} +\eta \tan ^{2} \theta)}}}
 \end{eqnarray}
From the above expressions, if we consider a FIIP we recover the dispersion
relation for Alfv\'en waves with a background flow (Carbonell et al.  2009), and if we remove the flow, the well-known
dispersion relation for Alfv\'en waves is recovered. Since the flow speed is much smaller than the Alfv\'en speed, the effect of the flow on the real
and imaginary parts of the wavenumber is very small, then, the wavelengths and damping lengths are similar to those 
in~\ref{Anf}. The third
remaining wavenumber of Eq. (\ref{disp_alf5}) can be approximated by, 
\be k = \frac{\omega}{v_\mathrm{0} \cos \theta} + i
\frac{(v_\mathrm{0}^{2} -v_\mathrm{a}^{2})\cos \theta}{v_\mathrm{0}(\eta_\mathrm{C}\cos^{2} \theta +\eta \sin^{2} \theta)}, \label{disp_alf4} \ee 
corresponding to the third Alfv\'en wave. All the above analytical approximations display an excellent agreement with the numerical results, and the presence of the third Alfv\'en wave, given by
Eq.~(\ref{disp_alf4}), fully depends on the join presence of flow and resistivities since, otherwise, the dispersion relation (\ref{disp_alf5})
would be quadratic.  For an external observer to the flowing plasma, this additional wave could be detectable, although its
strong spatial damping would make its detection very difficult. For an observer linked to the flow inertial rest frame,
only the two usual Alfv\'en waves, modified by resistivities, would be detected.

%%%%%%%%%%%%%%%%%%%%%%%%%%%%%%%%%%%%%%%%%
  \section{Spatial damping of Magnetoacoustic waves in a partially ionised 
    plasma}
%%%% %%%%%%%%%%%%%%%%%%%%%%%%%%%%%%%%%%%%%
    Our general dispersion relation for non-adiabatic magnetoacoustic waves in presence of a background flow is given by 
    Eq.~(\ref{disp_mag}). Because of the complexity of this dispersion relation, and in order to be able to understand the 
    results, we
    have split our study in a sequence of four different cases having dispersion relations of increasing complexity. In the first two cases,  the behaviour of adiabatic magnetoacoustic waves in a non-flowing~(\ref{anf}) and 
    flowing plasma~(\ref{awf}) is considered; while in the last two cases, the behaviour of non-adiabatic waves in a non-flowing~(\ref{Nanf}) and flowing plasma~(\ref{Nawf}) is studied.
 
    %%%%%%%%%%%%%%%%%%%%%%%%%%%%%
    \subsection{Adiabatic Magnetoacoustic waves without 
    background flow} \label{anf}
 %%%%%%%%%%%%%%%%%%%%%%%%%%%%%%  
    
Setting $A = H = 0$ in Eq.~(\ref{nass}), $v_\mathrm{0} = 0$ in expression~(\ref{om}) and substituting in Eq.~(\ref{disp_mag}), we
obtain the dispersion relation for adiabatic magnetoacoustic waves in a PIP without a background flow,
which is
\begin{eqnarray}
     (\omega^{2} -k^{2} c_\mathrm{s}^{2}) (ik^{2} \eta_\mathrm{C} \omega-\omega^{2})+k^{2} v_\mathrm{a}^{2}(\omega^{2} -k_{x}^{2} c_\mathrm{s}^{2} 
    ) + i k^{2} k_{z}^{2}v_\mathrm{a}^{2} c_\mathrm{s}^{2} \Xi \rho_{0} \omega= 0  \label{mgp}
    \end{eqnarray}
    already found by Forteza et al.  (2007)

    %%%%%%%%%%%%%%%%%%%%%%%%%%%%%%%
    \subsubsection{Fully ionised resistive plasma} \label{fir0}
    %%%%%%%%%%%%%%%%%%%%%%%%%%%%%%%%
   Now, imposing FIRP conditions from~\ref{le}, the dispersion relation~(\ref{mgp})
    simplifies to,
    \begin{eqnarray}
     (\omega^{2} -k^{2} c_\mathrm{s}^{2}) (ik^{2} \eta \omega-\omega^{2})+k^{2} v_\mathrm{a}^{2}(\omega^{2} -k_{x}^{2} c_\mathrm{s}^{2} 
    ) = 0 \label{mg0}
    \end{eqnarray}
    which once expanded gives a fourth degree polynomial in the wavenumber $k$,
    \begin{eqnarray}
    k^{4} c_\mathrm{s}^{2} (\cos ^{2} \theta  v_\mathrm{a}^{2}+i \omega \eta) 
    -k^{2} \omega^{2}(c_\mathrm{s}^{2}+v_\mathrm{a}^{2} + i \omega \eta) + 
    \omega^{4} = 0. \label{mg1}
    \end{eqnarray}
    Furthermore, when only longitudinal propagation is allowed ($\theta = 0$), 
  dispersion relation~(\ref{mg0}) can be factorized as
    \begin{eqnarray}
    (\omega^{2} - k^{2} c_\mathrm{s}^{2}) \left[\omega^{2} -k^{2}(v_\mathrm{a}^{2}+i \omega 
    \eta)\right] = 0 \label{mg11}
    \end{eqnarray}
    giving place to two undamped slow waves with a dispersion relation
    given by,
    $$k^{2} = \frac{\omega^{2}}{c_\mathrm{s}^{2}}$$
    and, since for longitudinal propagation fast waves become Alfv\'en waves, we obtain two Alfv\'en waves, damped by 
    resistivity, whose dispersion relation is,
    $$ k^{2} =  \frac{\omega^{2}}{v_\mathrm{a}^{2} + i \eta \omega}$$
   For these Alfv\'en waves,
    the real $(k_{r})$ and imaginary $(k_{i})$ parts of the wavenumber $k$ can be obtained from expressions~(\ref{kra}) 
    and~(\ref{kia}) setting $\theta = 0$. Then, in the case of a FIRP, when oblique propagation is not allowed, resistivity 
 does not affect slow waves, which propagate undamped, and only 
 affects fast waves.  Next, if we allow
 oblique propagation ($0 < \theta < \pi/2$) in dispersion relation~(\ref{mg1}), the 
 solutions for the wavenumbers are given by, 
    \begin{eqnarray}
     k_\mathrm{m}^{2} =  \frac{ \omega^2 (v_\mathrm{a}^{2} + c_\mathrm{s}^{2} +i \eta \omega) }{2 
     i c_\mathrm{s}^{2}(\eta \omega - i v_\mathrm{a}^{2} \cos^{2} \theta)} +
     \nonumber \\
       + \frac{ \omega^2 \sqrt{(v_\mathrm{a}^{2} + c_\mathrm{s}^{2}+i \eta \omega)^{2} - 4 
     i c_\mathrm{s}^{2}(\eta \omega-i v_\mathrm{a}^{2}  \cos^{2} \theta)}}{2 i 
     c_\mathrm{s}^{2}(\eta \omega - i v_\mathrm{a}^{2} 
  \cos^{2} \theta)} 
     \end{eqnarray}
     and
     \begin{eqnarray}
    k_\mathrm{n}^{2}  =  \frac{ \omega^2 (v_\mathrm{a}^{2} + c_\mathrm{s}^{2} +i \eta \omega) }{2 
     i c_\mathrm{s}^{2}(\eta \omega - i v_\mathrm{a}^{2} \cos^{2} \theta)} -
     \nonumber \\
     - \frac{ \omega^2 \sqrt{(v_\mathrm{a}^{2} + c_\mathrm{s}^{2}+i \eta \omega)^{2} - 4 
     i c_\mathrm{s}^{2}(\eta \omega-i v_\mathrm{a}^{2}  \cos^{2} \theta)}}{2 i c_\mathrm{s}^{2}(\eta\omega - i v_\mathrm{a}^{2} 
  \cos^{2} \theta)} 
     \end{eqnarray}
where $k_\mathrm{m}^{2}$ and $k_\mathrm{n}^{2}$ are the wavenumbers corresponding to two coupled and damped fast and slow waves propagating in
opposite directions.  Figure~\ref{f08} displays the behaviour of damping length, wavelength and the ratio of
damping length to wavelength versus period for fast and slow waves, respectively.  In these figures it is clearly observed
that the behaviour of the wavelength is the same for a FIRP and a FIIP, and that the only difference occurs at periods below
$10^{-6}$ s due to resistivity.  However, looking to the ratio of the damping length to the wavelength, we can conclude
that in a FIRP, and within the interval of periods of interest, the spatial damping of both waves is negligible.

 %%%%%%%%%%%%%%%%%%%%%%%%%%%%%%%
    \subsubsection{Partially ionised plasma} \label{pip0}
    %%%%%%%%%%%%%%%%%%%%%%%%%%%%%%%%
The dispersion relation is given by Eq.~(\ref{mgp}), and considering only longitudinal propagation
we obtain, 
 \begin{eqnarray}
    (\omega^{2} - k^{2} c_\mathrm{s}^{2}) \left[\omega^{2} -k^{2}(v_\mathrm{a}^{2}+i \omega 
    \eta_\mathrm{C})\right] = 0 \label{mg111}
    \end{eqnarray}
This could suggest that the consideration of longitudinal propagation in a FIRP or in
a PIP leads to the same dispersion relation, however, there is an important difference, while for FIRP both resistivities
have the same numerical value, for PIP the numerical value of Cowling's resistivity is much greater than that of Spitzer's
resistivity.  Furthermore, for longitudinal propagation the slow waves are not influenced by Cowling's resistivity. Once expanded, Eq.~(\ref{mg111}) gives the following fourth degree polynomial in the wavenumber $k$,
   \begin{eqnarray}
    k^{4}i  c_\mathrm{s}^{2} \left[\eta_\mathrm{C} \omega -i v_\mathrm{a}^{2} \cos ^{2} \theta + 
    v_{a}^{2} \rho_{0} \omega \Xi (\cos^{2} 
    \theta-1)\right] - \nonumber  \\
     - k^{2} \omega^{2} (v_\mathrm{a}^{2 } + c_\mathrm{s}^{2} +i \eta_\mathrm{C}\omega) + 
    \omega^{4} = 0.\label{mg}
    \end{eqnarray}
     After solving this biquadratic dispersion relation, we obtain, 
    \begin{eqnarray}
	k_\mathrm{m}^{2} = \frac{A+\sqrt {B}}{C} \\
		k_\mathrm{n}^{2} = \frac{A-\sqrt{B}}{C} 
	\end{eqnarray}
	with
     \begin{eqnarray}
     A  & = & (v_\mathrm{a}^{2} + c_\mathrm{s}^{2} +i \eta_\mathrm{C} \omega) \nonumber \\
      B  & = & (v_\mathrm{a}^{2} + c_\mathrm{s}^{2}+i \eta_\mathrm{C} \omega)^{2} - \nonumber \\ 
      &   & -4 
     i c_\mathrm{s}^{2}\left[\eta_\mathrm{C} \omega-i v_\mathrm{a}^{2}  \cos^{2} \theta - 
     v_\mathrm{a}^{2} \rho_{0} \Xi (\cos^{2} \theta -1)\right] \nonumber \\
     C  & = & 2 
     i c_\mathrm{s}^{2}\left[\eta_\mathrm{C} \omega - i v_\mathrm{a}^{2} \cos^{2} \theta - 
     v_\mathrm{a}^{2} \rho_{0} \Xi (\cos^{2} \theta -1)\right] \nonumber
     \end{eqnarray}
     and the wavenumbers are,
     \begin{eqnarray}
   k_\mathrm{1} = \pm \sqrt{\omega^{2}  k_\mathrm{m}^{2}}  \\
   k_\mathrm{2} = \pm \sqrt{\omega^{2}  k_\mathrm{n}^{2}} 
     \end{eqnarray}
   
     \begin{figure}
	  \centering{
		 \resizebox{9cm}{!} {\includegraphics{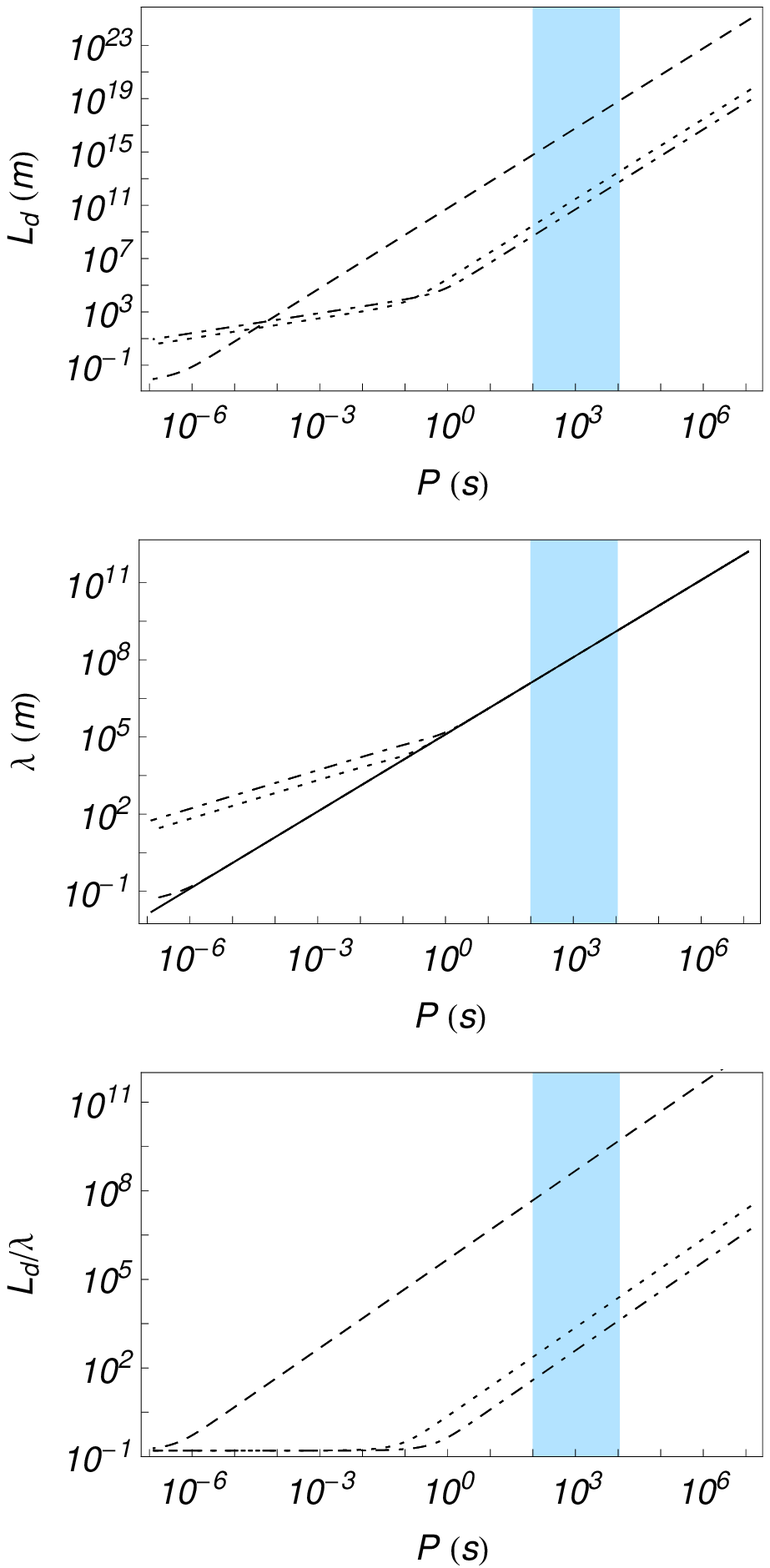}
		      \includegraphics{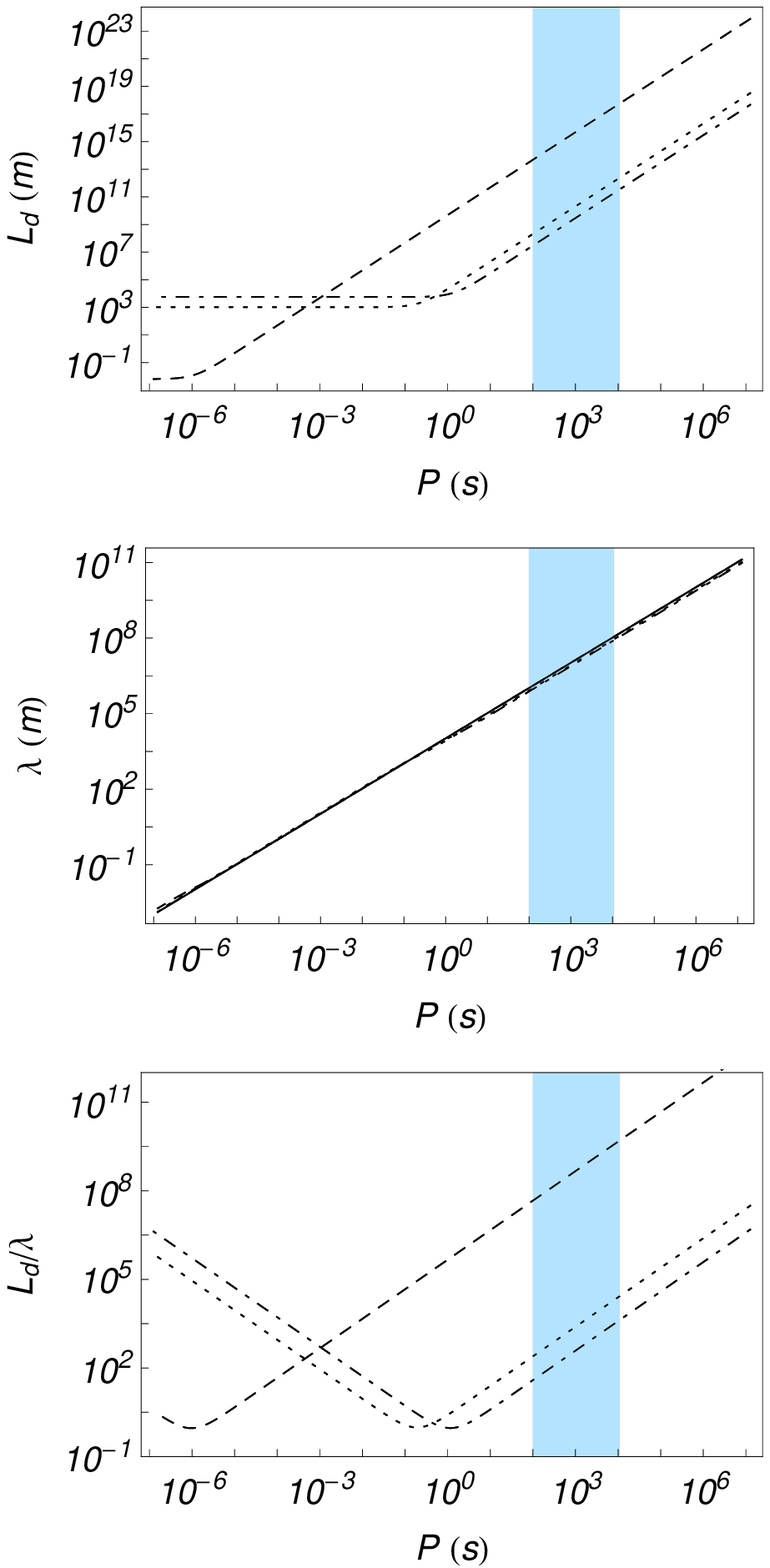}}}
		   \vspace{-3mm}
		   \caption{Damping length, wavelength, and ratio of the damping length to the wavelength versus period
for the adiabatic fast  (left panels) and slow (right panels) waves in a FIIP (solid), a FIRP
 (dashed) and a PIP with $\tilde{\mu}=0.8$ (dotted) and $\tilde{\mu}=0.95$ (dash-dotted).}   \label{f08} 
	   \end{figure}
	   
	   \begin{figure}
		  \centering{
		 \resizebox{9cm}{!} {\includegraphics{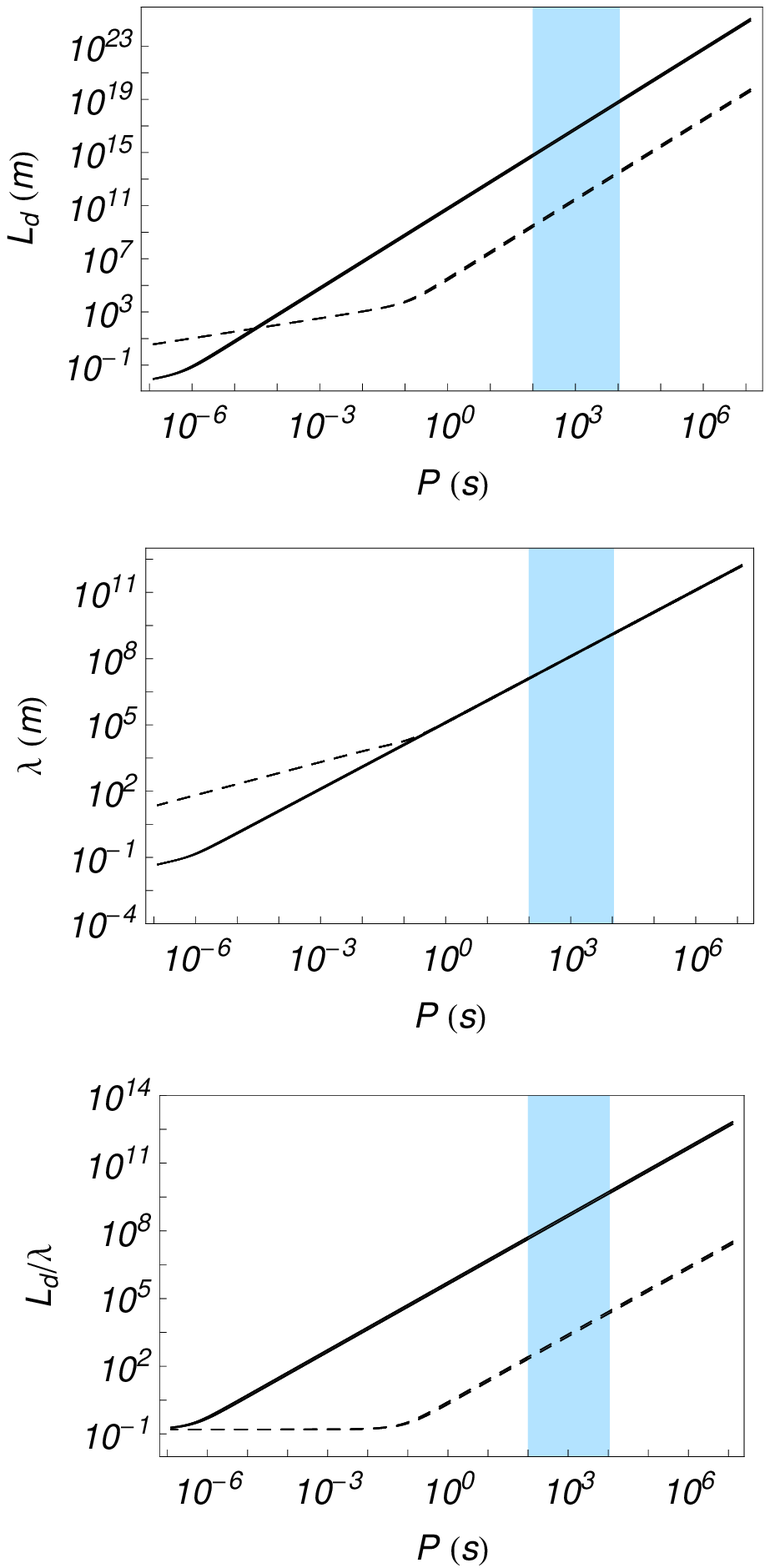}
		      \includegraphics{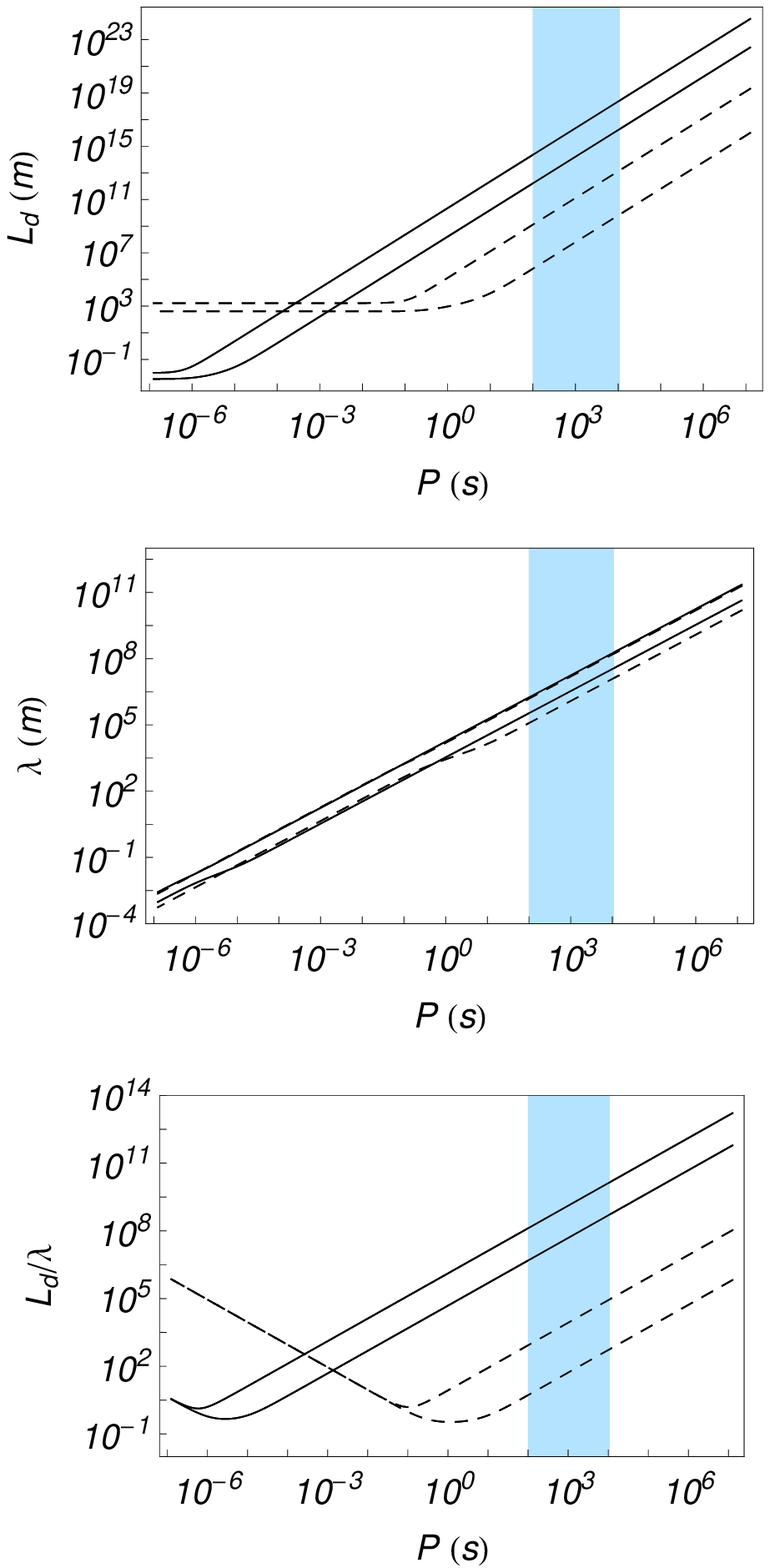}}}
 \vspace{-3mm}
		   \caption{Damping length, wavelength, and ratio of the damping length to the wavelength versus period
for the adiabatic fast (left panels) and slow (right panels) waves in a FIIP  (solid) and in a PIP with $\tilde{\mu}=0.8$ (dashed). The background flow speed is $10$ \ km/s. }   \label{f10} 
	   \end{figure}

	    \begin{figure}
	 	  \centering{
		 \resizebox{9cm}{!} {\includegraphics{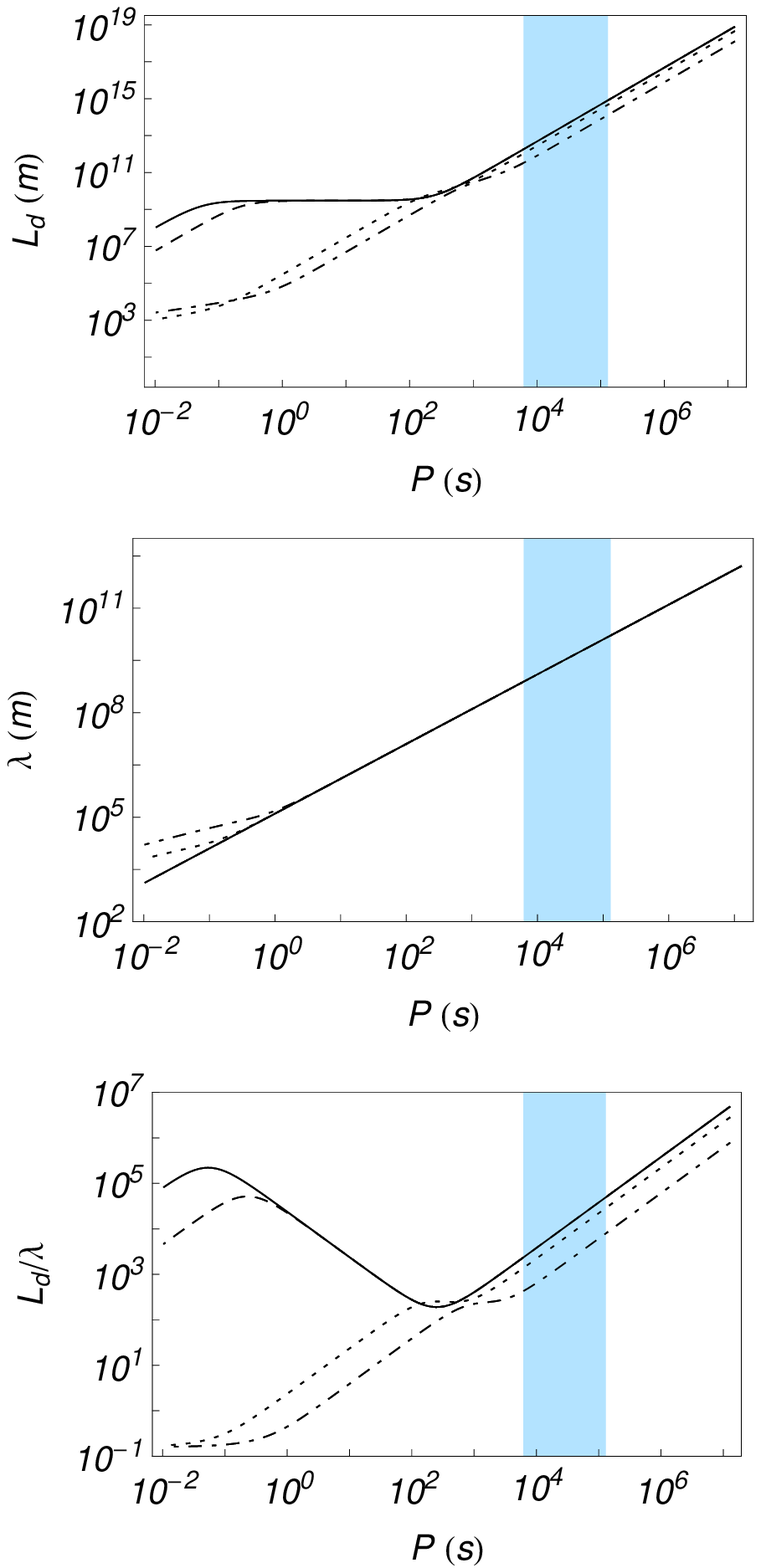}
		      \includegraphics{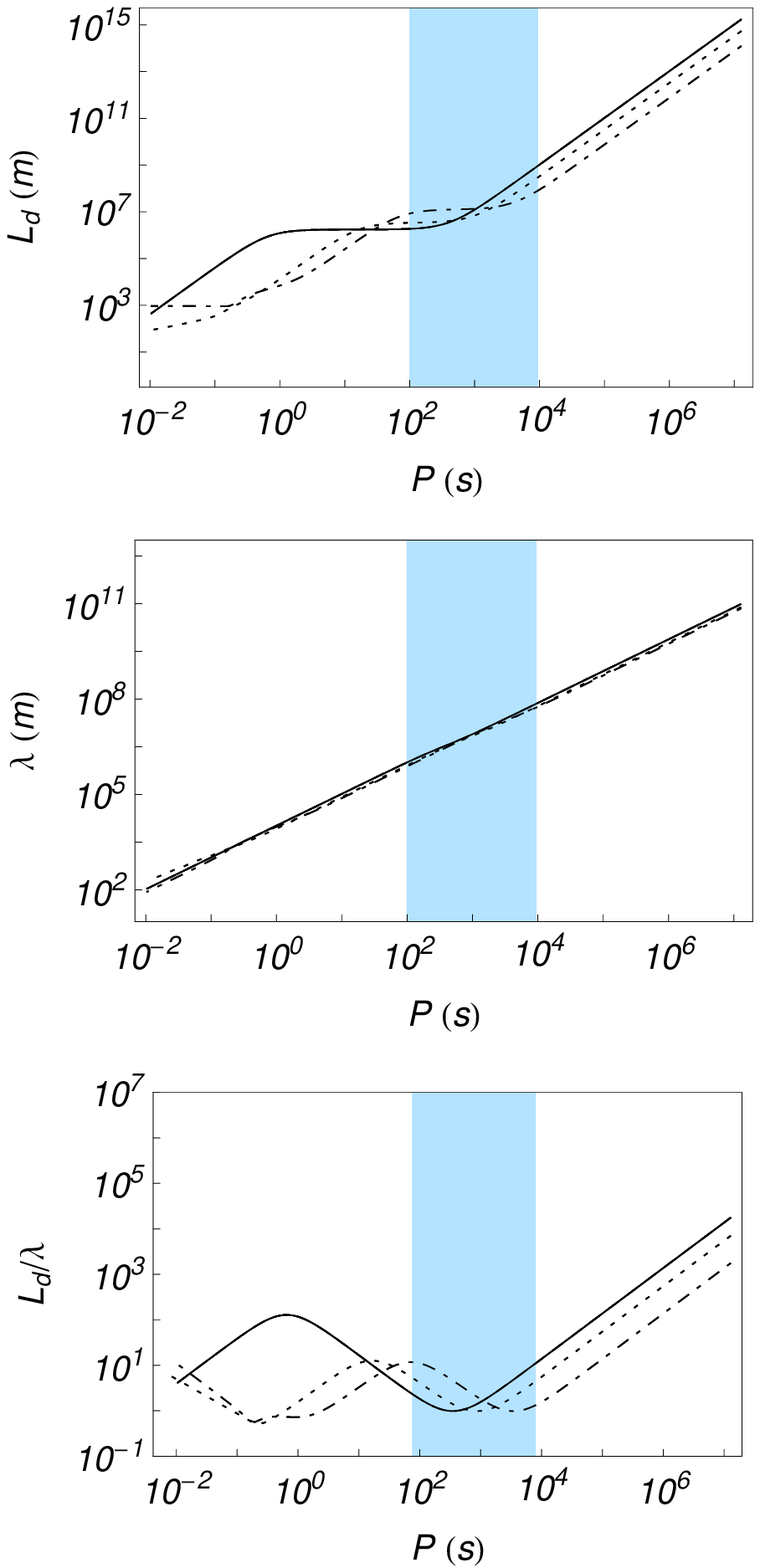}  }}
 \vspace{-3mm}
  \caption{Damping length, wavelength, and ratio of the damping length to the wavelength versus period
for the non-adiabatic fast (left panels), slow (right panels) waves in a FIIP (solid), a FIRP
(dashed), and a PIP with $\tilde{\mu}=0.8$ (dotted) and $\tilde{\mu}=0.95$ (dash-dotted). }   \label{f16} 
	   \end{figure}
	   
	     \begin{figure}
	 	  \centering{
		 \resizebox{9cm}{!}  {\includegraphics{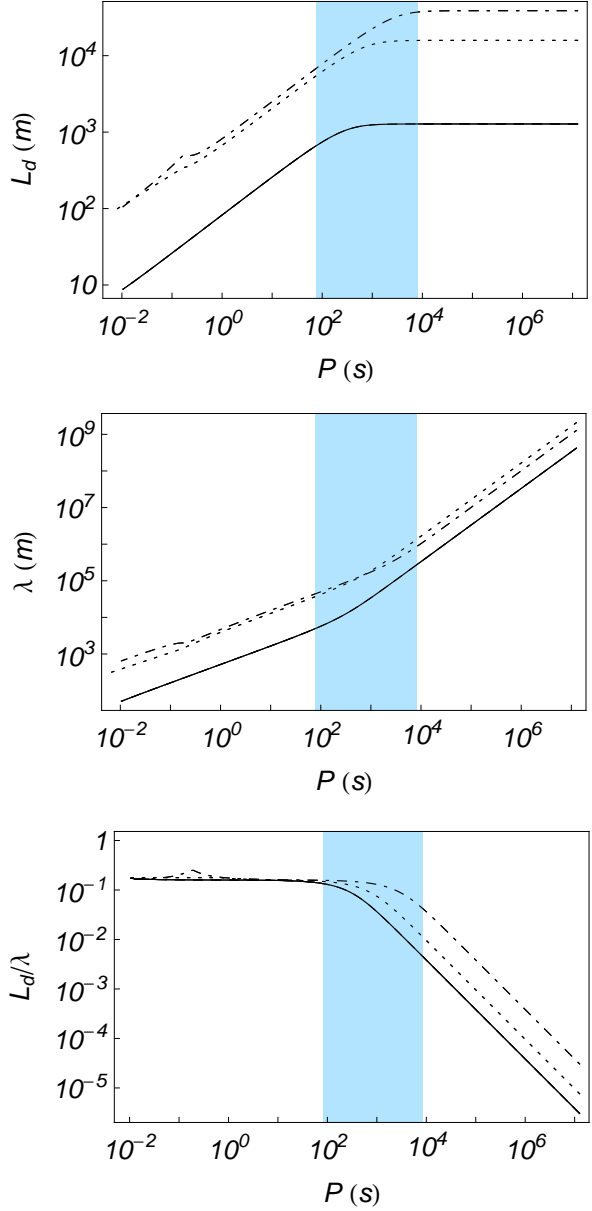} }}
 \vspace{-3mm}
 \caption{Damping length, wavelength, and ratio of the damping length to the wavelength versus period
for the non-adiabatic thermal wave in a FIIP (solid), a FIRP
(dashed), and a PIP with $\tilde{\mu}=0.8$ (dotted) and $\tilde{\mu}=0.95$ (dash-dotted). }   \label{f16a} 
	   \end{figure}

	    \begin{figure}
	   \centering{
		 \resizebox{9cm}{!} {\includegraphics{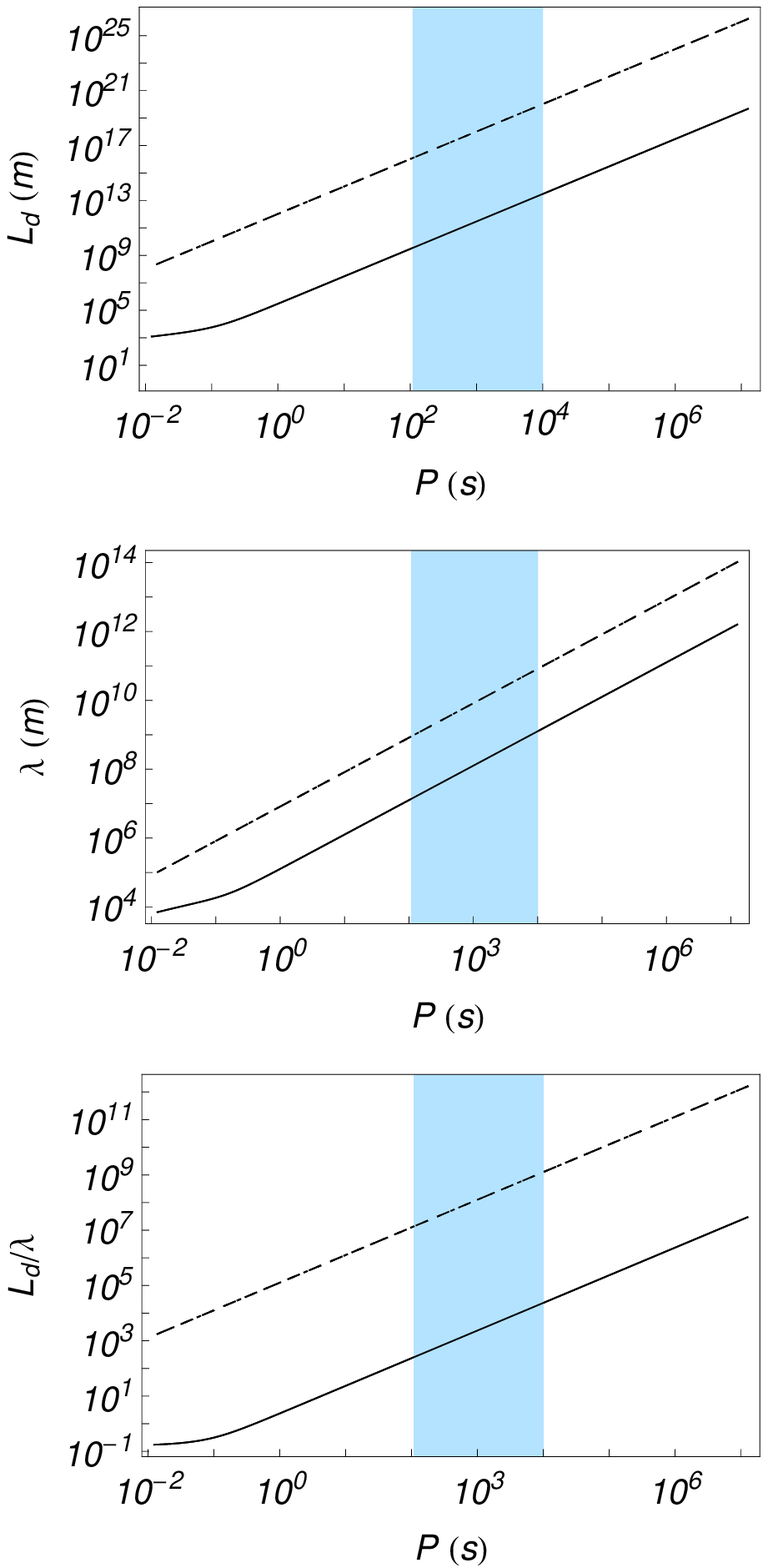}
		      \includegraphics{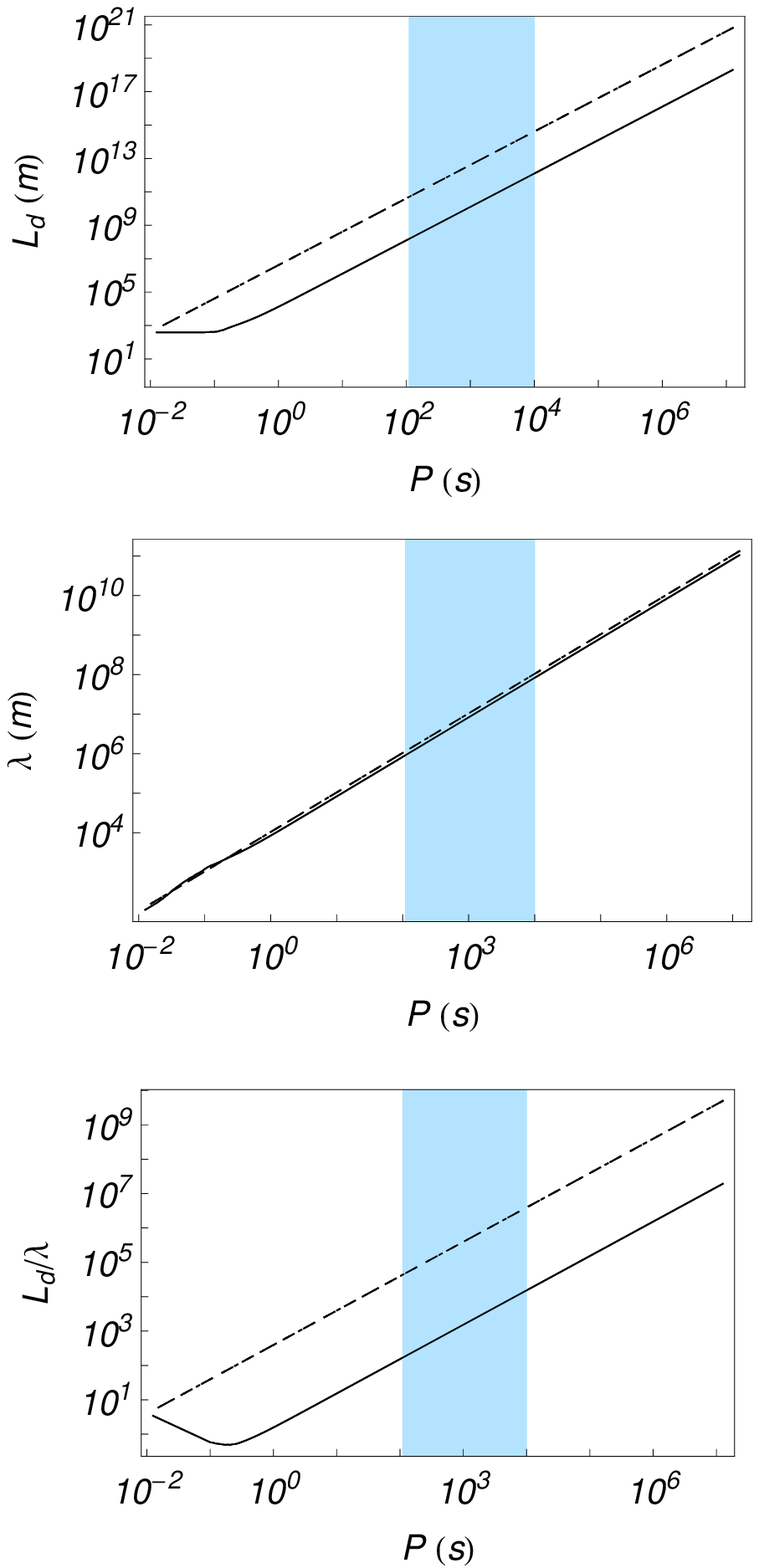}  }}
 \vspace{-3mm}
		   \caption{Damping length, wavelength, and ratio of the damping length to the wavelength versus period
for the non-adiabatic fast (left panels), slow (right panels) waves, without radiation, in a FIIP (dashed), a FIRP
(dotted) and a PIP with $\tilde{\mu}=0.8$ (solid).}   \label{f16c} 
	   \end{figure}
	   	   
 \begin{figure}
	   \centering{
		 \resizebox{9cm}{!} {\includegraphics{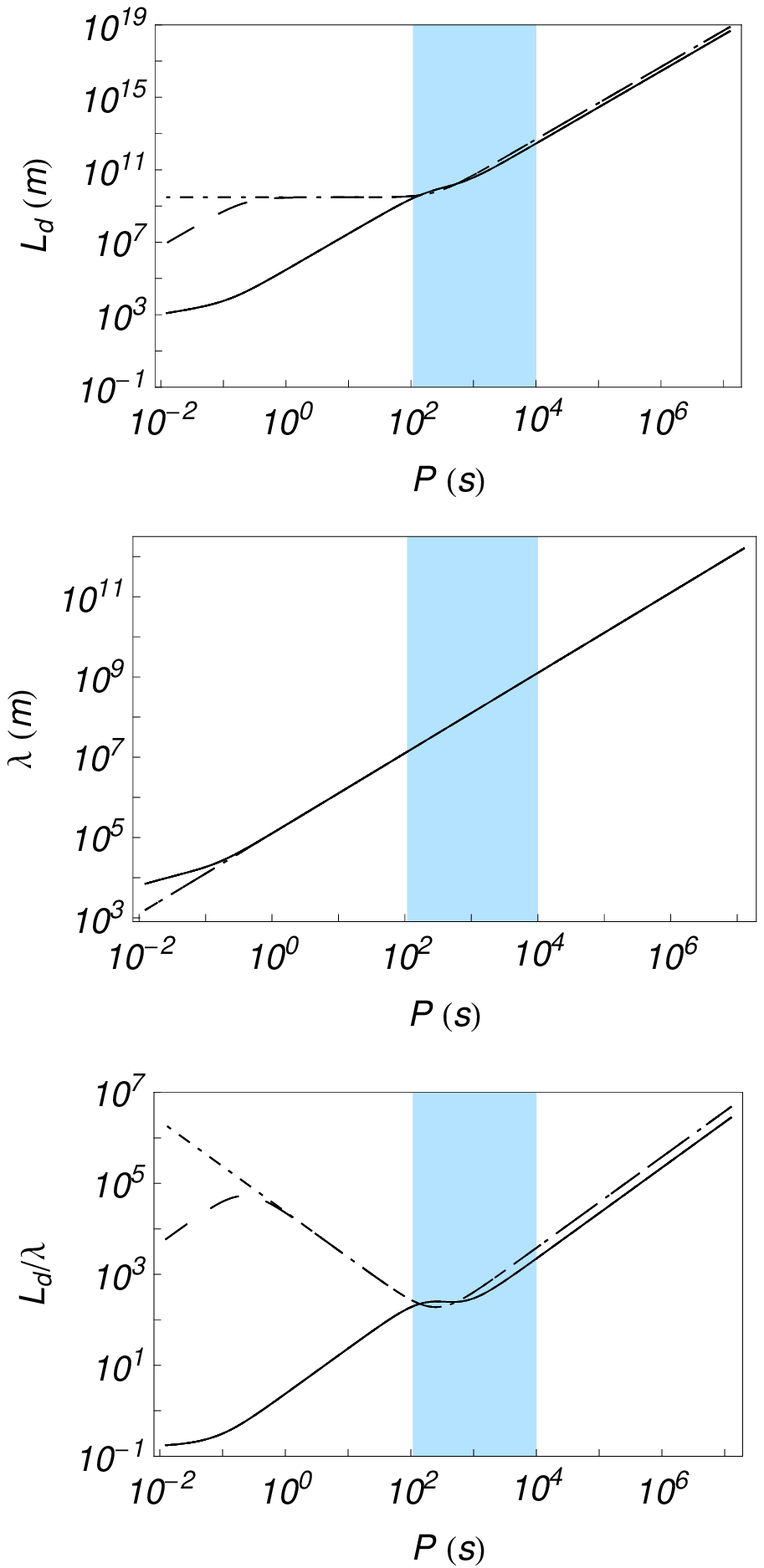}
		      \includegraphics{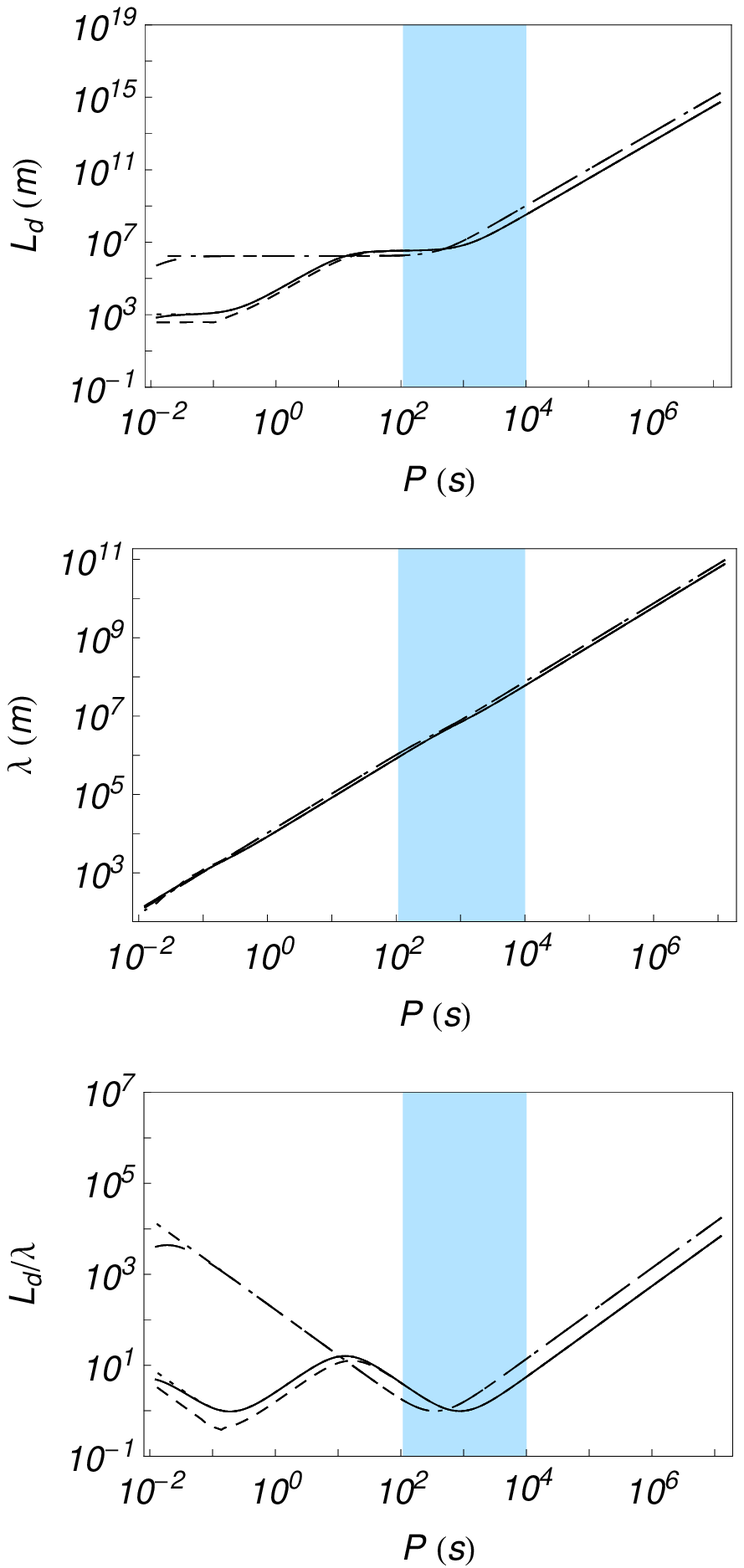} }}
 \vspace{-3mm}
		   \caption{Damping length, wavelength, and ratio of the damping length to the wavelength versus period
for the non-adiabatic fast (left panels), slow (right panels) waves in a PIP ($\tilde \mu = 0.8$) without neutrals thermal conduction (solid); 
without electronic thermal conduction (dashed) and without thermal conduction (dotted); in a FIIP without 
thermal conduction (dash-dotted), and in a FIRP without thermal conduction (long-dashed). }   \label{f16f} 
	   \end{figure}
	   
   \begin{figure}
	   \centering{
		 \resizebox{9cm}{!} {\includegraphics{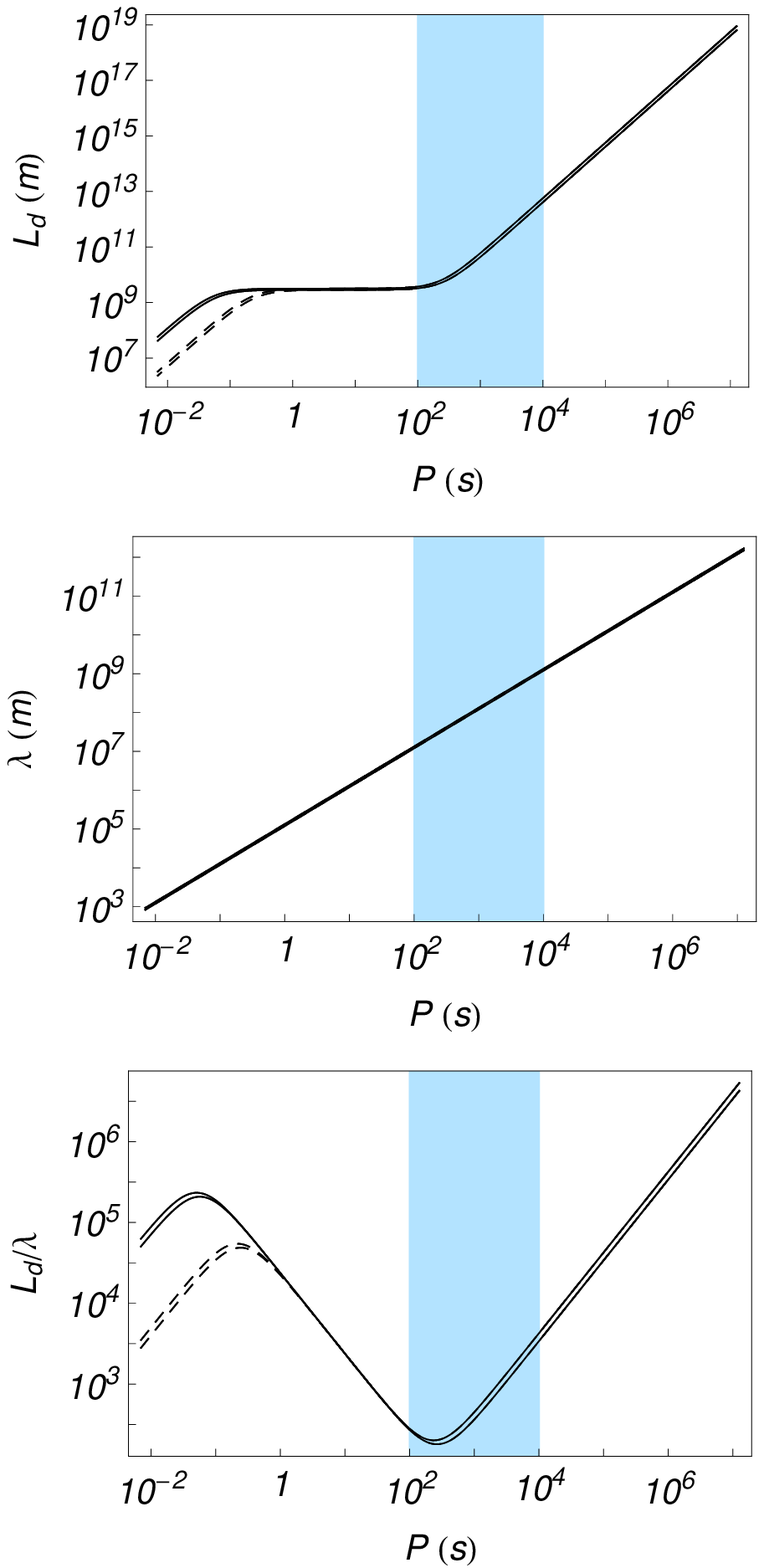}
		      \includegraphics{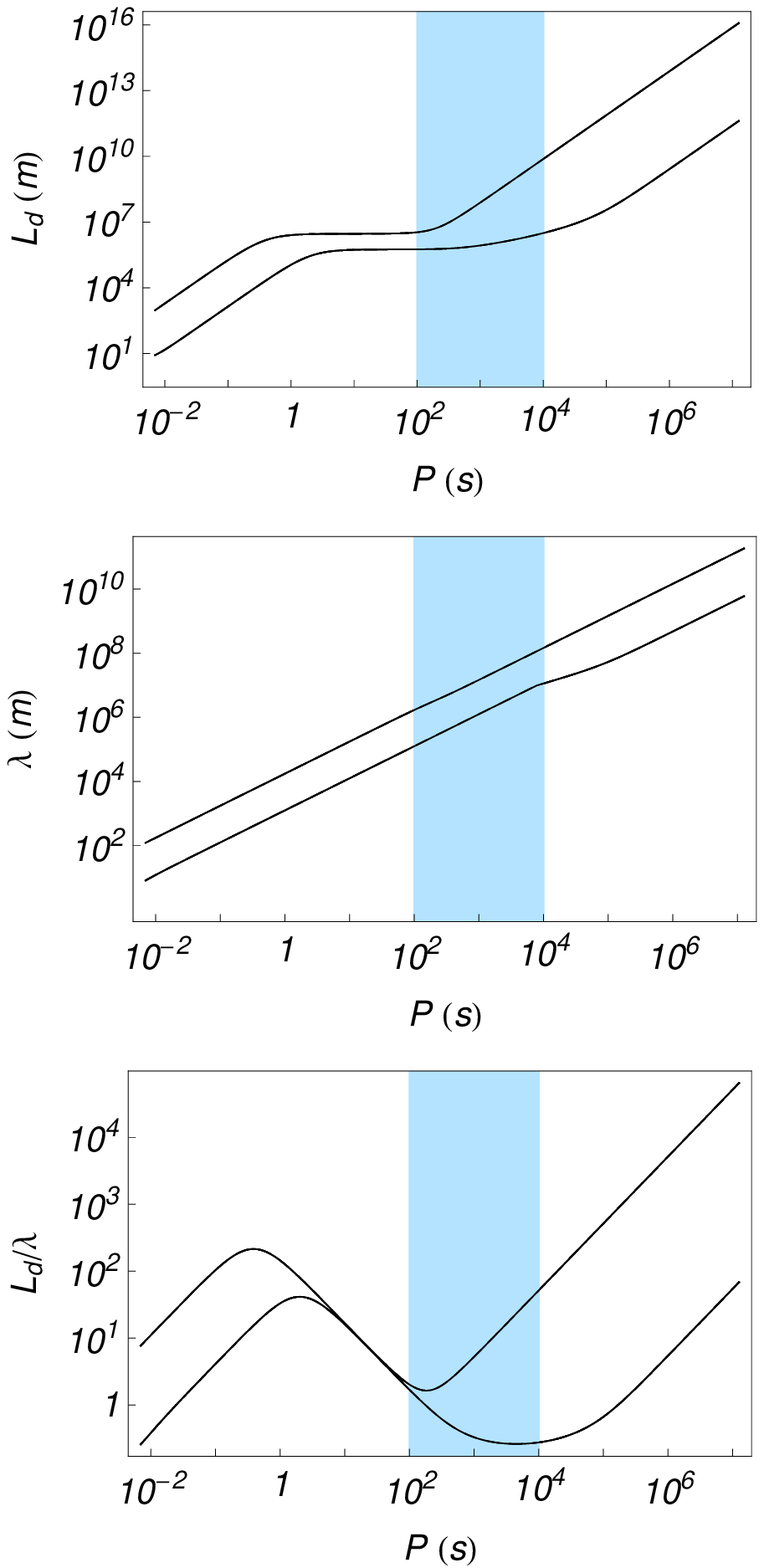} }}
 \vspace{-3mm}
		   \caption{Damping length, wavelength, and ratio of the damping length to the wavelength versus period
for the non-adiabatic fast (left panels), slow (right panels) waves in a FIIP (solid), and in a FIRP (dashed). The flow speed is $10$ \ km/s.}   \label{f19} 
	   \end{figure}

\noindent
	   Figure~\ref{f08} displays the behaviour of damping length, wavelength and the ratio of damping length to
wavelength versus period for fast and slow waves in a PIP, respectively.  The damping length of both slow and fast waves is severely
influenced by ion-neutral collisions showing a strong dependence on the period for periods greater than $1$ s while for
shorter periods the dependence becomes weaker.  This figure also shows that with respect to a FIIP the wavelength of the
fast waves is slightly affected by the partial ionisation, deviating from the linear behaviour for periods below $1$ s, while
the wavelength of slow waves is not affected at all.  Within the interval of observed periods in prominence oscillations,
when ionisation decreases the ratio between the damping length and the wavelength also decreases for both waves and the spatial damping
becomes more efficient. The maximum efficiency of the spatial damping for fast waves is
attained for periods below $1$ s while for slow waves the maximum of efficiency is attained at a period which depends
on the ionisation fraction.  The location of this maximum moves towards long periods when the ionisation of the plasma
decreases, but when almost neutral plasmas are considered it is still located at a period slightly greater than $1$ s, 
ouside the region of interest.  Finally,
comparing Figures~\ref{f01} and~\ref{f08} it becomes obvious that the behaviour of Alfv\'en and fast waves is quite similar. 
	   	   	 
 %%%%%%%%%%%%%%%%%%%%%%%%%%%%%%%%%%%%%%%%%%%%%    
 \subsection{Adiabatic Magnetoacoustic waves with background flow} \label{awf}
 %%%%%%%%%%%%%%%%%%%%%%%%%%%%%%%%%%%%%%%%%%%%%
 Setting $A = H = 0$ in Eq.~(\ref{nass}) and substituting in Eq.~(\ref{disp_mag}), we obtain the dispersion relation for adiabatic magnetoacoustic 
waves in a PIP with a background flow, which is
\begin{eqnarray}
     (\Omega^{2} -k^{2} c_\mathrm{s}^{2}) (ik^{2} \eta_\mathrm{C} \Omega-\Omega^{2})+k^{2} v_\mathrm{a}^{2}(\Omega^{2} -k_{x}^{2} c_\mathrm{s}^{2} 
    )+ \nonumber \\ 
    + i k^{2} k_{z}^{2}v_\mathrm{a}^{2} c_\mathrm{s}^{2} \Xi \rho_{0} \Omega= 0  \label{mgp1}
    \end{eqnarray}
 %%%%%%%%%%%%%%%%%%%%%%%%%%%%%%%
 \subsubsection{Fully ionised resistive plasma} \label{fir}
 %%%%%%%%%%%%%%%%%%%%%%%%%%%%%%%
 Considering FIRP conditions from~\ref{le}, the dispersion relation~(\ref{mgp1})
becomes, 
\begin{eqnarray}
(\Omega^{2} -k^{2} c_\mathrm{s}^{2})(ik^{2} \eta \Omega - \Omega^{2})+k^{2} v_\mathrm{a}^{2}(\Omega^{2}-k_{x}^{2}c_\mathrm{s}^{2})=0 \label{drmg}
\end{eqnarray}
The dispersion relation is a fifth degree polynomial in the wavenumber $k$, and for this reason we expect two slow waves and
two fast waves which, for the flow speed considered, propagate in opposite directions, plus an additional wave.  When only
longitudinal propagation is considered, the above dispersion relation becomes,
\begin{eqnarray}
(\Omega^{2} -k^{2} c_\mathrm{s}^{2})\left[k^{2} (i\eta \Omega +v_\mathrm{a}^{2})- \Omega^{2}\right]=0 \label{drmg1}
\end{eqnarray}
and slow waves are decoupled from fast waves propagating undamped, while fast waves are damped 
by resistivity. The wavenumbers corresponding to the undamped slow waves are given by, 
\begin{eqnarray}
    k = \frac{\omega}{v_\mathrm{0} \pm c_\mathrm{s}}
       \end{eqnarray}
Concerning the fast waves, now Alfv\'en waves because of longitudinal propagation, the corresponding dispersion relation, given by the second factor in Eq.~(\ref{drmg1}), is equivalent to 
Eq.~(\ref{disp_alf5}) when $\theta = 0$.  The solutions to this dispersion relation are given by Eqs.~(\ref{disp_alf3})
and~(\ref{disp_alf4}) with $\theta = 0$, and we obtain three Alfv\'en waves similar to those studied in \ref{Af}.  Then, the
expected additional wave mentioned above is a fast wave which, for longitudinal propagation and when a background flow is
present, becomes the third Alfv\'en wave already found in~\ref{Af}.  When oblique propagation is allowed, fast and slow waves
become coupled and the dispersion relation~(\ref{drmg}) is solved numerically.  Figure~\ref{f10} displays the
behaviour of the damping length, wavelength and the ratio of the damping length to wavelength versus period for fast and slow
waves in a FIRP.  Because of the strong difference between Alfv\'en and flow speeds, in the case of fast waves
the unfolding in wavelength and damping length due to the flow is not evident, while for slow waves it is
clearly seen.  In both cases the behaviour of the wavelength versus period is linear, similar to what
happens for a FIIP and the only difference is provided by the unfolding produced by the flow.  The damping length also
behaves linearly with period and, for slow waves, the unfolding in wavelength and damping length produces two different
curves for the ratio $L_\mathrm{d}/\lambda$.  In both curves, the most efficient spatial damping appears for periods far away
from those of interest in prominence oscillations.  For fast waves, the ratio $L_\mathrm{d}/\lambda$ behaves linearly with
period and its value is very large within the region of periods of interest.  The behaviour of the third fast wave is quite
different from the other two fast waves and very similar to that of the third Alfv\'en wave shown in  \ref{Af}, being strongly damped within the interval of periods considered.

  \begin{figure}
	  \centering{
		 \resizebox{9cm}{!} {\includegraphics{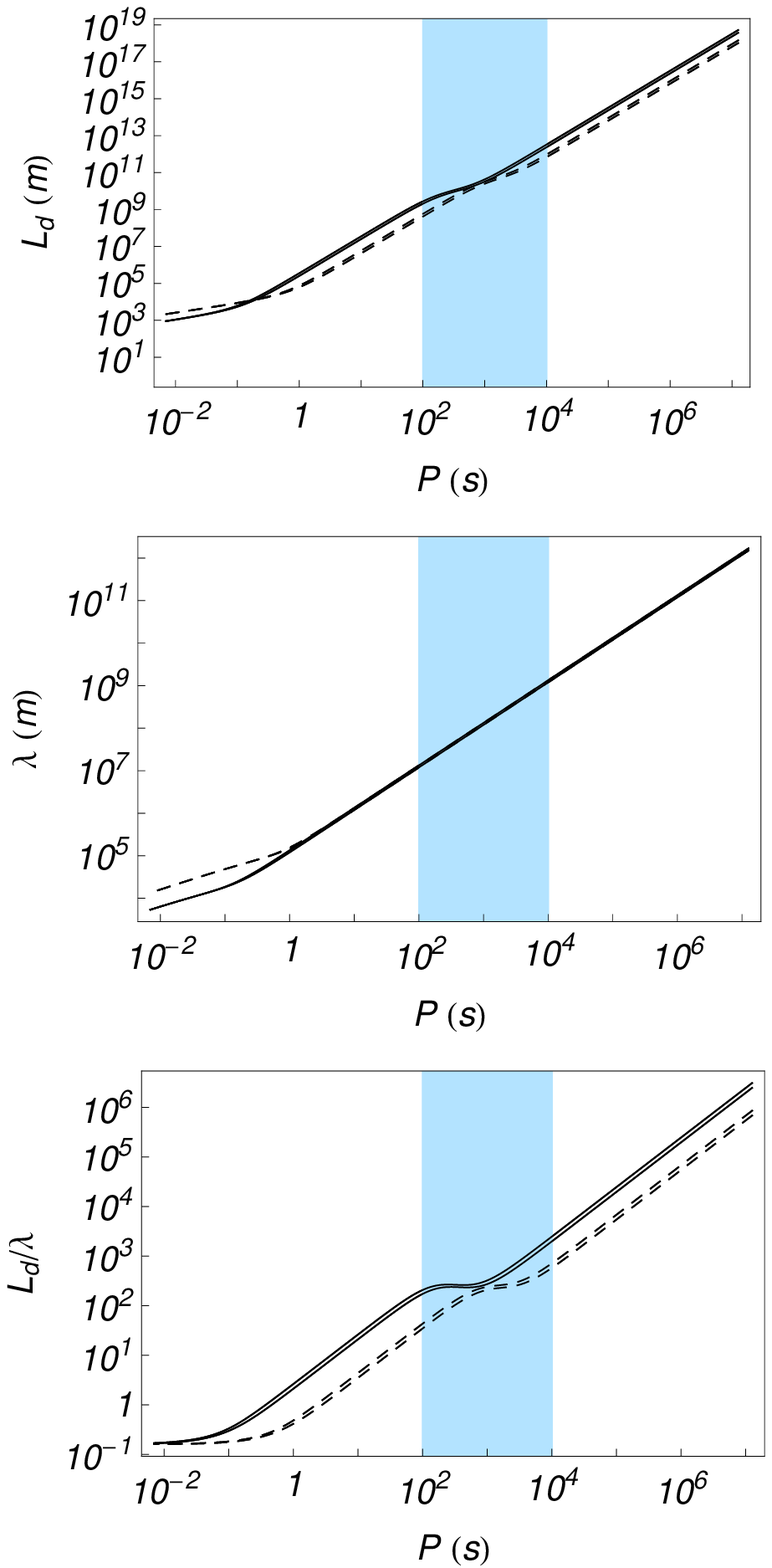}
		      \includegraphics{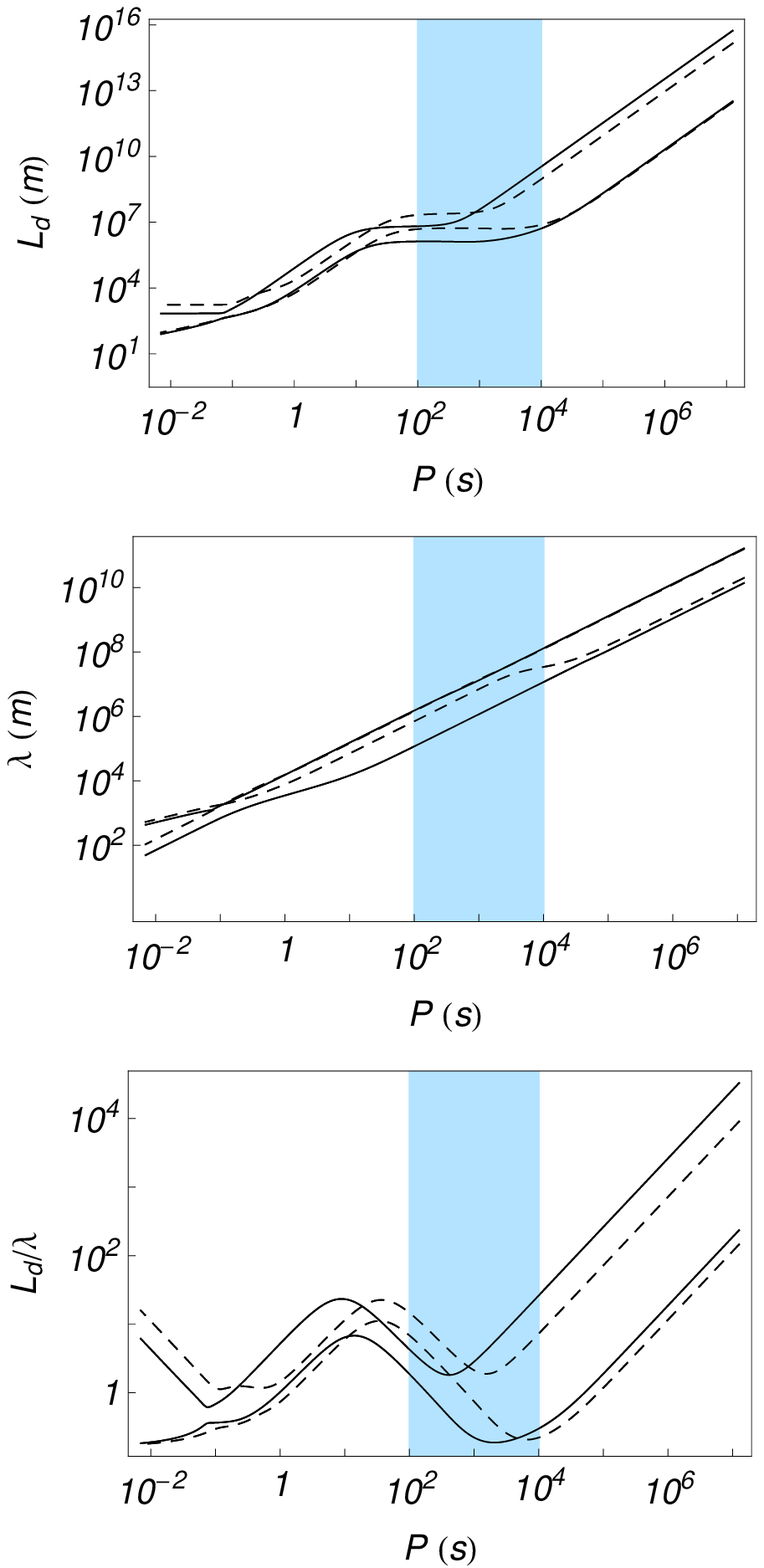}  }}
 \vspace{-3mm}
		   \caption{Damping length, wavelength, and ratio of the damping length to the wavelength versus period
for the non-adiabatic fast (left panels), slow (right panels) waves in a PIP with 
 $\tilde{\mu}=0.8$ (solid) and $\tilde{\mu}=0.95$ (dashed). The flow speed is $10$ \ km/s.}   \label{f23} 
	   \end{figure}

%%%%%%%%%%%%%%%%%%%%%%%%%%%%
\subsubsection{Partially ionised plasma}  \label{pip}
%%%%%%%%%%%%%%%%%%%%%%%%%%%%%
Our dispersion relation is given by Eq.~(\ref{mgp1}) which is a fifth degree polynomial in $k$, and when longitudinal
propagation is considered we recover the results of~\ref{fir} with slight differences due to the different numerical value of
Cowling's resistivity.  Such as it is shown in Figure~\ref{f10}, at periods longer than $0.1$ s, the damping
lengths of both slow and fast waves increase linearly with the period.  However, for periods below $0.1$ s the damping length
slowly decreases in the case of fast waves and becomes constant for slow waves. Furthermore, the wavelength of fast waves
for periods below $0.1$ s increases with respect to the FIRP case while the wavelengths corresponding to slow waves are only
slightly modified.  Concerning the ratio of the damping length versus the wavelength, the unfolding in wavelength caused by
the flow and the change of the damping length due to the partial ionisation produces that fast waves are much more
efficiently attenuated, than in a a FIRP, for any period, but especially at periods below $0.1$ s, while for slow waves the peak of maximum
efficiency is displaced towards long periods when ionisation decreases, and for almost neutral plasmas would approach the
region of periods usually observed in prominence oscillations. Again, the behaviour of the remaining third fast wave is very similar to that found for the
 third Alfv\'en wave discussed in \ref{Af}.

In absence of flow the dispersion relation~(\ref{drmg}) becomes a fourth
degree polynomial in $k$ and the third fast wave disappears.  Such as it happens for Alfv\'en waves with a background flow,
this third fast wave appears due to the joint action of flow and resistivity, since when a FIIP with a background flow is
considered, the dispersion relation also becomes a fourth degree polynomial in $k$ and the third fast wave is absent.

%%%%%%%%%%%%%%%%%%%%%%%%%%%%%%%%%%%%%%%%%%%%%%%
\subsection{Non-adiabatic Magnetoacoustic waves in a prominence plasma without background flow} \label{Nanf}
%%%%%%%%%%%%%%%%%%%%%%%%%%%%%%%%%%%%%%%%%%%%%%%
Setting $v_\mathrm{0} = 0$ in expression~(\ref{om}) and substituting in Eq.~(\ref{disp_mag}), we obtain the dispersion relation for non-adiabatic magnetoacoustic 
waves in a PIP without a background flow, which is
\begin{eqnarray}
     (\omega^{2} -k^{2} \Lambda^{2}) (ik^{2} \eta_\mathrm{C} \omega-\omega^{2})+k^{2} v_\mathrm{a}^{2}(\omega^{2} -k_{x}^{2} \Lambda^{2}
    )+ \nonumber \\ 
    + i k^{2} k_{z}^{2}v_\mathrm{a}^{2} \Lambda^{2} \Xi \rho_{0} \omega= 0  \label{mgp2}
    \end{eqnarray}

%%%%%%%%%%%%%%%%%%%%%%%%%%
 \subsubsection{Fully ionised resistive plasma} \label{Nafir}
 %%%%%%%%%%%%%%%%%%%%%%%%%%%
 Once imposed the conditions corresponding to a FIRP, the following dispersion relation, a sixth degree polynomial in the 
 wavenumber $k$, is obtained,
 \begin{eqnarray}
(\omega^{2} -k^{2} \Lambda^{2})(ik^{2} \eta \omega - \omega^{2})+k^{2} v_\mathrm{a}^{2}(\omega^{2}-k_{x}^{2} \Lambda^{2})=0 
\label{drmg3}
\end{eqnarray}
which describes coupled fast, slow and thermal waves. When 
only longitudinal propagation is allowed, the above dispersion relation becomes,
\begin{eqnarray}
(\omega^{2} -k^{2} \Lambda^{2})\left[k^{2}(i \eta \omega +v_\mathrm{a}^{2})- \omega^{2}\right]=0 
\label{drmg4}
\end{eqnarray}
Then, like in~\ref{Anf} we obtain two decoupled Alfv\'en waves, damped by resistivity, whose dispersion relation is given by,
 \be k^2 = \frac{\omega^{2}}{v_\mathrm{a}^{2}+i\omega \eta} \ee 
 and another dispersion relation,
 \begin{eqnarray}
  k^{2} = \frac{\omega^{2}}{ \Lambda^{2}}  \label{ts}   
  \end{eqnarray}
which is a fourth degree polynomial in $k$ describing coupled propagating thermal and slow waves, only damped by thermal
 effects. The dispersion relation~(\ref{drmg3}) has been solved numerically and Figures~\ref{f16} and \ref{f16a}
 show the behaviour of the damping length, wavelength and ratio of the damping length to the wavelength versus period for the
 fast, slow and thermal waves. Starting with Figure~\ref{f16} we have only considered 
 an interval of periods between $10^{-2}$ and $10^{7}$ s, since for periods smaller than $10^{-2}$ s, much shorter than 
 those of interest, the curves become very entangled. In the case of the fast wave, when an FIIP is considered the spatial damping is
 governed by radiative losses and thermal conduction (Carbonell et al.  2006) in the interval of periods from $10^{-2}$ to
 $10^{7}$ s.  However, in a FIRP we observe a
 slight change in the damping length of fast waves around a period of $1$ s.  This change tells us that the dominance of
 thermal conduction appears slightly before than in the ideal case. Also, looking to the ratio between the damping length and the wavelength we observe that in
 this case the efficiency of the damping for fast waves is improved, with respect to a FIIP, for periods below
 $1$ s.  For slow waves, no differences appear in the behaviour of FIIP and FIRP. The behaviour  of thermal waves (Figure~\ref{f16a}) is exactly the same in FIIP and FIRP.
  
 %%%%%%%%%%%%%%%%%%%%%%
 \subsubsection{Partially ionised plasma}
 %%%%%%%%%%%%%%%%%%%%%
Now, our dispersion relation is given by Eq.~(\ref{mgp2}) and when only longitudinal propagation
is allowed the expression is similar to that of~\ref{Nafir}, although the numerical value of the Cowling's resistivity is
different and the results for fast waves would be different.  When oblique propagation is considered, the dispersion
relation~(\ref{mgp2}) has been solved numerically and a strong distortion of the damping length and wavelength curves
corresponding to fast waves (Figure~\ref{f16}, left panels) appears.  Basically, the changes affect the radiative plateau, between
periods $10^{3}$ and $10^{-2}$ s, and partial ionisation decreases the damping length
of fast waves in this region.  For slow waves (Figure~\ref{f16}, right panels), a similar behaviour is found in the same regions although the
distortion is not so important since a very short radiative plateau, between $10^{2}$ and $10^{3}$ s remains, together with 
a region, between $1$ and $10^{2}$ s, where thermal conduction is
dominant.  As compared to a FIRP, the ratio $L_\mathrm{d}/\lambda$
for fast waves decreases substantially for periods below $10^{3}$ s, although for the periods of interest in
prominences this ratio remains very large.  For slow waves, partial ionisation produces that the ratio $L_\mathrm{d}/\lambda$
reaches a maximum of efficiency, $\sim 1$, for periods similar to those involved in prominence oscillations, and this maximum is
displaced towards longer periods when ionisation is decreased.  The changes in the wavelengths of slow and fast waves are
similar to those shown in the adiabatic case (Sect.~\ref{pip0}). In the case of thermal waves (Figure~\ref{f16a}), partial ionisation increases both the damping length and wavelength of these waves, although the behaviour of the ratio $L_\mathrm{d}/\lambda$ is similar to previous cases. Since thermal wave is always strongly damped, which makes its detection very difficult, in the following we will avoid further comments on it.

In order to understand the effects of radiation and thermal conduction by neutrals and electrons, in Figure~\ref{f16c}
 we represent, for fast and slow waves, the same quantities but  with optically thin
radiation and heating removed i.e. only thermal conduction is at work, and we can observe that the most efficient
damping for both waves occurs for partially ionised plasmas, which suggests that the inclusion of the isotropic thermal
conduction due to neutrals plays a very important role for all the periods considered.  However, we must take into account
that when the ionisation fraction decreases, radiation also decreases and that, because of neutrals, thermal conduction is favoured, what makes
difficult to establish meaningful comparisons between plasmas with different degrees of ionisation.

Next, in Figure~\ref{f16f} we plot the behaviour of fast and slow waves,
 when thermal conduction has been removed i.e. only radiation and heating remain.  When thermal conduction by
electrons and neutrals is removed, the behaviour of fast waves in a FIIP and in FIRP is similar and, within the interval of
periods of interest, the behaviour of the different quantities corresponding to the different types of plasma is the same.
However, when periods below $100$ s are considered, FIIP and FIRP are strongly affected by the lack of thermal conduction.
For slow waves, a similar behaviour appears and the most important conclusion is that when we compare 
Figures~\ref{f16c} and \ref{f16f}, and within the interval of periods of interest, the presence of a very efficient damping is due to optically thin
radiation.

On the other hand, when non-adiabatic magnetoacoustic waves are considered, the importance of radiation and thermal
conduction can be also quantified in terms of two dimensionless parameters (De Moortel and Hood 2004), namely, the thermal
ratio, which modified for the case of partial ionisation becomes,
      
      \begin{equation}
            d = \frac {(\gamma -1) (\kappa_{\mathrm{e}\parallel} + \kappa_\mathrm{n})T_{0}
            \rho_{0}}{\gamma^2 p^2_{0} \tau_{\mathrm{s}}} =
            \frac{1}{\gamma}\frac{\tau_{\mathrm{s}}}{\tau_{\mathrm{cond}}},
         \end{equation}
      which is $1/\gamma$ times the ratio of the sound travel time
      $(\tau_{\mathrm{s}} = l/c_{\mathrm{s}})$ to the thermal
      conduction timescale $(\tau_{\mathrm{cond}} = l^2
      p_{0}/\left[(\gamma - 1) (\kappa_{\mathrm{e}\parallel}+\kappa_\mathrm{n}) T_{0}\right])$, and the
      radiation ratio,
      
      \begin{equation}
      r = \frac{(\gamma - 1) \tau_{\mathrm{s}} \xi_{i} \rho^2_{0} \chi^{*}
      T^{\alpha}_{0}}{\gamma p_{0}} = \frac
      {\tau_{\mathrm{s}}}{\tau_{\mathrm{r}}},
      \end{equation}
      which is the ratio of the sound travel time to the radiation timescale
      $(\tau_{\mathrm{r}} = \gamma p_{0}/\left [(\gamma - 1) \xi_{i} \rho^2_{0} \chi^{*}
      T^{\alpha}_{0}\right])$. From the equilibrium parameters, we can
      compute the value of
      $l$ at
      which the condition $d = r$ is satisfied,
      \begin{equation} \label{eqr}
          l = \sqrt{\frac
          {\kappa_{\mathrm{e}\parallel} + \kappa_\mathrm{n}
	  }{\xi_{i} \rho_{0}^2 \chi^{*}T^{\alpha-1}_{0}}}
             \end{equation}
where $\kappa_{e\parallel}$ and $\kappa_{n}$ are the thermal conduction coefficients corresponding to anisotropic electronic
and isotropic, by neutrals, thermal conduction, respectively.  Then, when the spatial length of the perturbation, the
wavelength, is of the order of $l$ or smaller, thermal conduction becomes dominant.  For the slow wave, Figure~\ref{f16} (Right panels)
shows that the transition from a regime dominated by radiation to another dominated by conduction can be clearly seen in the plot of the damping
length versus period.  For a FIIP, this transition occurs at a period of $1$ s while for a PIP, with $\tilde \mu = 0.8$, occurs at a
period between $10$ and $100$ s, and with $\tilde \mu = 0.95$, occurs at a period close to $100$ s.  Using the values assumed
for prominence parameters, for FIIP we obtain $l \approx 4700$~m; for a PIP ($\tilde \mu = 0.8$), $l \approx 40000$~m; and
for a PIP ($\tilde \mu = 0.95$), $l \approx 100000$~m.  Then, setting these periods in the plot of the wavelength versus period in Figure~\ref{f16} (Right panels), we can check that the numerical value of the wavelength is almost coincident with the above analytical
determinations for $l$.  The reason for this increase in the numerical value of $l$ is that for a PIP thermal conduction is
enhanced due to neutrals contribution, increasing when the ionisation fraction decreases, while the denominator of $l$
decreases when the ionisation fraction decreases.  Summarizing, when ionisation decreases the period at which the dominant
damping mechanism changes from radiation to thermal conduction increases and, consequently, the wavelength also increases.

  %%%%%%%%%%%%%%%%%%%%%%%%%%%%%%%%%%%%%%%%%%%%%%
  \subsection{Non-adiabatic Magnetoacoustic waves in a prominence plasma with background flow} \label{Nawf}
  %%%%%%%%%%%%%%%%%%%%%%%%%%%%%%%%%%%%%%%%%%%%%%
  
  %%%%%%%%%%%%%%%%%%%%
 \subsubsection{Fully ionised ideal plasma}
 %%%%%%%%%%%%%%%%%%%%
 Setting in Eq.~(\ref{disp_mag}) conditions corresponding to FIIP,  we obtain, 
  \begin{eqnarray}
(\Omega^{2} -k^{2} \Lambda^{2}) (-\Omega^{2})+k^{2} v_\mathrm{a}^{2}(\Omega^{2}-k_{x}^{2} \Lambda^{2})=0 
\label{drmg5}
 \end{eqnarray}
 which is a sixth degree polynomial in the wavenumber $k$.  When only longitudinal propagation is allowed, we obtain two 
 undamped Alfv\'en waves given by, 
 \be k^{2} = \frac{\Omega^{2}}{v_\mathrm{a}^{2}} \ee
 whose solutions for the wavenumbers are
 \be k = \frac{\omega}{v_\mathrm{0} \pm v_\mathrm{a}} \ee
 as in Carbonell et al. (2009), while the dispersion relation,
 \be \Omega^{2} = k^{2} \Lambda^{2} \ee
 describes coupled slow and thermal waves modified by the flow and damped by thermal effects. In the case of oblique propagation, dispersion 
 relation~(\ref{drmg5}) has been solved numerically and 
 the behaviour of fast and slow waves is shown in Figure~\ref{f19}. When a flow is present the unfolding of wavelengths and damping lengths appears.  Since the considered 
 flow speed is much smaller than Alfv\'en speed, the separation of
 the curves corresponding to the fast wave is very small while, since flow speed and sound speed are comparable, the curves
 corresponding to slow waves substantially separate.  This effect strongly affects the behaviour of the damping
 length versus wavelength for slow waves, since one of them has a very efficient spatial damping for periods observed in
 prominence oscillations. In the case considered in this section, the damping of fast and slow waves is strictly due to  thermal effects.

 %%%%%%%%%%%%%%%%%%%%%%%%%
  \subsubsection{Fully ionised resistive plasma}
  %%%%%%%%%%%%%%%%%%%%%%%%%%
  In this case, considering FIRP conditions, the dispersion relation becomes
    \begin{eqnarray}
(\Omega^{2} -k^{2} \Lambda^{2}) (ik^{2} \eta  \Omega-\Omega^{2})+k^{2} v_\mathrm{a}^{2}(\Omega^{2}-k_{x}^{2} \Lambda^{2})=0 
\label{drmg6}
 \end{eqnarray}
which is a seventh degree polynomial in the wavenumber $k$.  Considering only longitudinal propagation, we find, again, coupled
slow and thermal waves modified by the flow and damped by thermal effects, and three Alfv\'en waves given by the dispersion
relation~(\ref{disp_alf5}), with $\theta = 0$, and its solutions.  Therefore, when we solve dispersion
relation~(\ref{drmg6}), we expect three fast, two slow and two thermal propagating waves.  In Figure~\ref{f19}
fast and slow waves have been plotted and compared with the previous case.  Such as it is shown, for
fast and slow waves no important differences in the behaviour with respect to the ideal case are seen.
Thermal waves, as well as the third
fast wave, are strongly damped after a very short distance.

%%%%%%%%%%%%%%%%%%%%%
  \subsubsection{Partially ionised plasma}
  %%%%%%%%%%%%%%%%%%%%% 
Now, the dispersion relation is given by Eq.~(\ref{disp_mag}) and Figure~\ref{f23} displays¼ the
behaviour of the damping length, wavelength and ratio of damping length versus wavelength for fast and slow  waves.
The most interesting results are those related with the ratio $L_\mathrm{d}/\lambda$.  For fast waves,
this ratio decreases with the period becoming small for periods below $10^{-2}$ s while for one of the slow waves,
 the ratio becomes very small for periods typically observed in prominence oscillations.  When ionisation
is decreased, slight changes of the above described behaviour occur, the most important being the displacement towards longer
periods of the peak of most efficient damping corresponding to slow waves. Such as it has been pointed out before, the presence of a third fast wave is, again, due to the joint action of flow and 
resistivity. In absence of flow or resistivity, the dispersion relation would become a  sixth order polynomial in the wavenumber and 
this wave would be absent.

%%%%%%%%%%%%%%%%%%%
\section{Conclusions}
%%%%%%%%%%%%%%%%%%%

Quiescent solar prominences and filaments are partially ionised plasmas and some of their typical features are the presence of
material flows and oscillations.  While the time damping of these oscillations has been thoroughly studied, their spatial
damping demands an in-depth study.  Interpreting the observed oscillations in terms of MHD waves, we have analysed the
spatial damping of Alfv\'en and non-adiabatic magnetoacoustic waves in a flowing partially ionised prominence plasma.
Several different cases, with dispersion relations of increasing complexity, have been considered, and the conclusions derived from our study are summarised in the following.

As it is well known, Alfv\'en waves are difficult to damp since non-adiabatic effects do not affect them, however, when
Alfv\'en waves in a partially ionised plasma are considered, they can be spatially damped and analytical
expressions describing their spatial damping can be obtained.  When the ionisation decreases, the damping length of these waves
also decreases and the efficiency of their spatial damping in the range of periods of interest is improved, although the most
efficient damping is attained for periods below $1$ s.  A new feature is that when a flow is present a
new third Alfv\'en wave, strongly attenuated, appears.  The presence of this wave depends on the joint action of flow and
resistivities, since in absence of flow, or for a FIIP, the dispersion relation becomes quadratic giving
place to the two well-known Alfv\'en waves. Furthermore, this third wave could only be detected by an observer not moving with the flow.

When adiabatic magnetoacoustic waves are considered and the effect of partial ionisation is taken into account, some new
features appear.  When a FIRP is considered and only longitudinal propagation is allowed, slow waves are decoupled from fast
waves, propagating undamped while fast waves propagate with a modified Alfv\'en speed and are damped by resistivity.
When a PIP plasma is studied and only longitudinal propagation is allowed, the same happens but now the numerical value of
Cowling's resistivity is greater than before enhancing the damping. Then,  the behaviour  of slow waves is only influenced by partial ionisation
when oblique propagation is allowed.  Furthermore, when a PIP is considered the behaviour of fast waves is very similar
to that of Alfv\'en waves and the damping becomes very efficient for periods below $1$ s, while for slow waves the peak
denoting the most efficient damping moves towards higher periods when the plasma ionisation decreases.  When in the adiabatic
case a flow is considered, the main difference is the unfolding of the damping length, wavelength and damping length to
wavelength curves, and the apparition of a third fast wave, strongly damped,  due to the join presence of flow and resistivities.  

Considering now non-adiabatic magnetoacoustic waves, when partial ionisation is present the behaviour of fast, slow and
thermal waves is strongly modified.  Comparing with non-adiabatic fast waves in a FIIP, which are damped by
electronic thermal conduction and radiation, the damping length of a fast wave in a PIP is strongly diminished by neutrals
thermal conduction for periods between $0.01$ and $100$ s, and, at the same time, the radiative plateau present in FIIP and
FIRP disappears. The behaviour of slow waves is not so strongly modified as for fast waves, although thermal conduction
by neutrals also diminishes the damping length for periods below $10$ s, and a short radiative plateau still remains for
periods between $10$ and $1000$ s.  Finally, thermal waves are only slightly modified although the effect of partial
ionisation is to increase the damping length of these waves, just the opposite to what happens with the other waves.  Next,
when a background flow is included, a new third fast wave appears which, again, is due to the joint action of flow and
resistivities.  As we already know, wavelengths and damping lengths are modified by the
flow, and since for slow waves sound speed and observed flow speeds are comparable this means that the change in wavelength
and damping length are important leading to an improvement in the efficiency of the damping.  Also, the maximum of efficiency
is displaced towards long periods when the ionisation decreases, and for ionisation fractions from $0.8$ to $0.95$ it is clearly
located within the range of periods typically observed in prominence oscillations with a value of $L_\mathrm{d}/\lambda$ smaller
than $1$.  This means that for a typical period of $10^{3}$ s, the damping length would be between $10^{2}$ and $10^{3}$ km,
the wavelength around $10^{3}$ km and, as a consequence, in a distance smaller than a wavelength the slow wave would be
strongly attenuated. On the other hand, during our calculations we have found that the different heating mechanisms usually 
considered (Carbonell et al. 2004) do not affect the results.

In conclusion, the joint effect of non-adiabaticity, flows and partial ionisation allows  slow waves  to damp in an efficient
way within the interval of periods typically observed in prominences.  Thermal waves are attenuated very efficiently within
the interval of interest but, probably, their observational detection is very difficult, and fast waves are very unefficiently
atenuated within the considered interval.  Furthermore, since Alfv\'en and fast waves display a similar behaviour, the
observational discrimination between them should be based on the detection of their associated perturbations.  It is also
important to remark that the new fast and Alfv\'en waves are only detectable in a reference system external to the flow,
however, its short damping length should make its detection very difficult.  Such as we have seen, even in the most simple
case of an unbounded medium threaded by a uniform magnetic field, the inclusion of non-adiabatic effects, partial ionisation
and flows complicates the study of the spatial damping of prominence oscillations because of the apparition of new waves and
the difficulty to discriminate between the different effects.  This also points out that from observations, and due to the
entanglement of the different effects, it is extremely difficult to properly interpret the observed oscillations in
terms of MHD waves.

The present study represents a first step in the investigation of the behaviour of the spatial damping of MHD waves in 
partially ionised prominence plasmas. Forthcoming studies must focus in investigating the behaviour of this type of 
damping in fine structures, threads, which we think are the basic constituents of quiescent prominences.

{\bf Acknowledgements}

The authors acknowledge the financial support provided by MICINN and
FEDER funds under grant AYA2006-07637. P. Forteza acknowledges a  FPU fellowship from MECyT. Also, the Conselleria
d'Economia, Hisenda i Innovaci\'o of the Government of the Balearic
Islands is gratefully acknowledged for the funding provided under grant
PCTIB2005GC3-03.

%%%%%%%%% References  %%%%%%

\end{document}